\documentclass[twocolumn]{aastex631}
\usepackage{longtable}
\usepackage[T1]{fontenc}
\usepackage{mathtools}

\defcitealias{2023AA...672A..93M}{MES-23}
\defcitealias{2023AA...672A..94D}{DER-23}
\defcitealias{2017AA...602A.107B}{BON-17}
\defcitealias{2010AA...524A..55G}{GAN-10}
\defcitealias{2011AA...528A..97T}{TIN-11}
\defcitealias{2012AA...537A.136G}{GUE-12}
\defcitealias{2012AA...545A...6P}{PAT-12}
\defcitealias{2013AA...555A.118A}{ALM-13}
\defcitealias{2022MNRAS.516..636S}{SEB-22}
\defcitealias{2008AA...488L..47M}{MOU-08}
\defcitealias{2009AA...506..281R}{RAU-09}
\defcitealias{2010AA...512A..14F}{FRI-10}
\defcitealias{2020NatAs...4..408H}{HAS-20}
\defcitealias{2018AJ....156....3C}{CHA-18}
\defcitealias{2021MNRAS.505.4956B}{BEN-21}
\defcitealias{2018AJ....156..213M}{MEN-18}
\defcitealias{2011AJ....142...95K}{KIP-11}
\defcitealias{2012AJ....144...19B}{BAK-12}
\defcitealias{2012AJ....144..139H}{HAR-12}
\defcitealias{2014AJ....147..128H}{HAR-14}
\defcitealias{2014AJ....147...84B}{BIE-14}
\defcitealias{2015AJ....150..168H}{HAR-15}
\defcitealias{2015AJ....150...85H}{HUA-15}
\defcitealias{2021AJ....162....7B}{BAK-21}
\defcitealias{2008ApJ...673L..79N}{NOY-08}
\defcitealias{2016AJ....152..182H}{HAR-16}
\defcitealias{2019AJ....158..141Z}{ZHO-19}
\defcitealias{2019AJ....157...82W}{WAN-19}
\defcitealias{2016AJ....152...88R}{RAB-16}
\defcitealias{2019RNAAS...3...35O}{OLI-19}
\defcitealias{2016AJ....152..108E}{ESP-16}
\defcitealias{2013AJ....146..113B}{BAY-13}
\defcitealias{2016AJ....152..161D}{DEV-16}
\defcitealias{2018AJ....155..119B}{BAY-18}
\defcitealias{2018MNRAS.477.3406B}{BEN-18}
\defcitealias{2018AJ....155..112B}{BRA-18}
\defcitealias{2018AJ....155...79H}{HEN-18}
\defcitealias{2019AJ....158...63E}{ESP-19}
\defcitealias{2019AJ....157...55H}{HAR-19}
\defcitealias{2022AA...660A.124S}{SRE-22}
\defcitealias{2013AA...551A..90M}{MAR-13}
\defcitealias{2006ApJ...646..505B}{BUT-06}
\defcitealias{2019AJ....158..138S}{STA-19}
\defcitealias{2017AA...599A..57B}{BOR-17}
\defcitealias{2022AA...667L..14Z}{ZAK-22}
\defcitealias{2004AA...415..391M}{MAY-04}
\defcitealias{2017AJ....153..136S}{STA-17}
\defcitealias{2016AA...588A.145H}{HEB-16}
\defcitealias{2009AA...496..513M}{MOU-09}
\defcitealias{2010AA...523A..15N}{NAE-10}
\defcitealias{2021AA...651A..11D}{DAL-21}
\defcitealias{2006ApJ...637.1094F}{FIS-06}
\defcitealias{2010ApJ...720.1290G}{GHE-10}
\defcitealias{2014MNRAS.438.2413V}{VON-14}
\defcitealias{2018AJ....155..136M}{MAY-18}
\defcitealias{2008ApJ...677.1324T}{TOR-08}
\defcitealias{2012ApJ...756L..33Q}{QUI-12}
\defcitealias{Pont_2011}{PON-11}
\defcitealias{2016AA...591A.146D}{DIA-16}
\defcitealias{2015ApJ...806....5H}{HAR-15}
\defcitealias{2009AA...503..601B}{BAR-09}
\defcitealias{2021ApJS..255....8R}{ROS-21}
\defcitealias{2018AA...615A.175B}{BAR-18}
\defcitealias{2008AJ....136.1901L}{LOP-08}
\defcitealias{2021AA...653A..78D}{DEM-21}
\defcitealias{2007AA...470..721N}{NAE-07}
\defcitealias{2017MNRAS.466..443J}{JEN-17}
\defcitealias{2006ApJ...647..600J}{JOH-06}
\defcitealias{2019ApJS..242...25F}{FEN-19}
\defcitealias{2020AJ....159..241D}{DAL-20}
\defcitealias{2005AA...444L..21G}{GAL-05}
\defcitealias{2003AA...410.1039P}{PER-03}
\defcitealias{2007ApJ...670.1391R}{ROB-07}
\defcitealias{2009ApJ...703.1545F}{FIS-09}
\defcitealias{2017AJ....153..130E}{EIG-17}
\defcitealias{2023AA...677A..33B}{BON-23}
\defcitealias{2018MNRAS.477.2572B}{BRA-18-2}
\defcitealias{2018MNRAS.481..596J}{JOH-18}
\defcitealias{2016PASP..128l4402B}{BRA-16}
\defcitealias{2016AJ....152..193B}{BAR-16}
\defcitealias{2012ApJ...761..123S}{SIV-12}
\defcitealias{2017AJ....153..178S}{STE-17}
\defcitealias{2019AA...625A..16J}{JON-19}
\defcitealias{2018AJ....155..100J}{JOH-18-2}
\defcitealias{2019AJ....158..197R}{ROD-19}
\defcitealias{2015AJ....150...12B}{BIE-15}
\defcitealias{2019AJ....157..192C}{CHO-19}
\defcitealias{Kipping_2022}{KIP-22}
\defcitealias{2011AA...533A..83B}{BOU-11}
\defcitealias{2011AA...528A..63S}{SAN-11}
\defcitealias{2015AA...575A..71A}{ALM-15}
\defcitealias{2010ApJ...713L.131K}{KOC-10}
\defcitealias{2013AA...554A.114H}{HEB-13}
\defcitealias{2010ApJ...724.1108J}{JEN-10}
\defcitealias{2017AA...606A..73T}{TAL-17}
\defcitealias{2018MNRAS.481.4960R}{RAY-18}
\defcitealias{2023MNRAS.518.4845J}{JAC-23}
\defcitealias{2020MNRAS.491.2834C}{COS-20}
\defcitealias{2008AA...482..299U}{UDA-08}
\defcitealias{2019AJ....157..224A}{ALS-19}
\defcitealias{2017AJ....153..200A}{ALS-17}
\defcitealias{2019AJ....157...74A}{ALS-19-2}
\defcitealias{2022AA...664A..94P}{PSA-22}
\defcitealias{2019MNRAS.490.1094K}{KOS-19}
\defcitealias{2021AJ....162..218C}{CAB-21}
\defcitealias{2021AJ....161..235H}{HOB-21}
\defcitealias{2023MNRAS.521.2765R}{ROD-23}
\defcitealias{2022AJ....164...70Y}{YEE-22}
\defcitealias{2021MNRAS.502.3704A}{ADD-21}
\defcitealias{2021ApJ...920L..16D}{DON-21}
\defcitealias{2023ApJS..265....1Y}{YEE-23}
\defcitealias{Brahm_2023}{BRA-23}
\defcitealias{2023AJ....165..121H}{HEI-23}
\defcitealias{2023AA...672L...7K}{KHA-23}
\defcitealias{2022AA...666A..46U}{ULM-22}
\defcitealias{2020AJ....160..209I}{IKW-20}
\defcitealias{2022AJ....163....9I}{IKW-22}
\defcitealias{2021AJ....161..194R}{ROD-21}
\defcitealias{2020AJ....159..145J}{JOR-20}
\defcitealias{2020AJ....160..235B}{BRA-20}
\defcitealias{2014MNRAS.440.1982H}{HEL-14}
\defcitealias{2016ApJ...823...29A}{ADD-16}
\defcitealias{2014AA...570A..64S}{SMI-14}
\defcitealias{2014AA...568A..81L}{LEN-14}
\defcitealias{2016MNRAS.463.3276H}{HAY-16}
\defcitealias{2016PASP..128f4401T}{TUR-16}
\defcitealias{2016MNRAS.458.4025D}{DEL-16}
\defcitealias{2016AA...591A..55M}{MAX-16}
\defcitealias{2017MNRAS.465.3693H}{HEL-17}
\defcitealias{2009AJ....137.4834W}{WES-09}
\defcitealias{2020AJ....159..255C}{COO-20}
\defcitealias{2019MNRAS.482.1379H}{HEL-19}
\defcitealias{2019MNRAS.489.2478N}{NIE-19}
\defcitealias{2011MNRAS.416.2108A}{AND-11}
\defcitealias{2019MNRAS.488.3067H}{HEL-19-2}
\defcitealias{2020AA...633A..30M}{MAN-20}
\defcitealias{2019MNRAS.490.1479H}{HEL-19-3}
\defcitealias{2020MNRAS.499..428S}{SCH-20}
\defcitealias{2019AJ....157..141T}{TEM-19}
\defcitealias{2010AJ....140.2007M}{MAX-10}
\defcitealias{2015AA...575A..61A}{AND-15}
\defcitealias{2012MNRAS.423.1503B}{BRO-12}
\defcitealias{2010PASP..122.1465M}{MAX-10-2}
\defcitealias{2011AA...525A..54B}{BAR-11}
\defcitealias{2015AA...577A..54C}{CIC-15}
\defcitealias{2017MNRAS.464..810B}{BRO-17}
\defcitealias{2009ApJ...690L..89H}{HEL-09}
\defcitealias{2013AA...552A.120S}{SMI-13}
\defcitealias{2013AA...552A..82G}{GIL-13}
\defcitealias{2014AA...563A.143D}{DEL-14}
\defcitealias{2016AA...585A.126W}{WES-16}
\defcitealias{2012MNRAS.427.1877C}{CAP-12}
\defcitealias{2008ApJ...683.1076W}{WIN-08}
\defcitealias{2017AJ....153...94C}{CRO-17}
\defcitealias{2020AJ....159...44C}{CRO-20}
\defcitealias{2023ApJ...950L..19F}{FRA-23}
\defcitealias{2008AA...487..373S}{SOU-08}
\defcitealias{2023RAA....23e5022X}{XIA-23}
\defcitealias{2007ApJ...669.1336F}{FIS-07}
\defcitealias{2016ApJ...825...53H}{HIR-16}
\defcitealias{2015AA...575A..85B}{BON-15}
\defcitealias{2011AA...528A..63S}{SAN-11}
\defcitealias{2019ApJ...877L..29C}{CAN-19}
\defcitealias{2010AA...524A..25T}{TRI-10}

\begin{document}

\title{Statistics and Habitability of F-type Star--Planet Systems}

\author[0000-0003-3920-7853]{Shaan D. Patel}
\author[0000-0002-8883-2930]{Manfred Cuntz}
\author[0000-0001-9194-2084]{Nevin N. Weinberg}
\affiliation{Department of Physics, University of Texas at Arlington, Arlington, TX 76019, USA}



\begin{abstract}
F-type star--planet systems represent an intriguing case for habitability studies.  Although F-type stars spend considerably less time on the main-sequence than G, K, and M-type stars, they still offer a unique set of features, allowing for the principal possibility of exolife.  Examples of the latter include the increased widths of stellar habitable zones as well as the presence of enhanced UV flux, which in moderation may have added to the origin of life in the Universe. In this study, we pursue a detailed statistical analysis of the currently known planet-hosting F-type stars by making use of the NASA Exoplanet Archive.  After disregarding systems with little or no information on the planet(s), we identify 206 systems of interest.  We also evaluate whether the stars are on the main-sequence based on various criteria. In one approach, we use the stellar evolution code \texttt{MESA}.  Depending on the adopted criterion, about 60 to 80 stars have been identified as main-sequence stars.  In 18 systems, the planet spends at least part of its orbit within the stellar habitable zone.  In one case, i.e., HD 111998, commonly known as 38 Vir, the planet is situated in the habitable zone at all times.  Our work may serve as a basis for future studies, including studies on the existence of Earth-mass planets in F-type systems, as well as investigations of possibly habitable exomoons hosted by exo-Jupiters as the lowest-mass habitable zone planet currently identified has a mass estimate of 143 Earth masses.
\end{abstract}

\keywords{Astrobiology; Catalogs; Exoplanets; Stellar types; Stellar physics; Stellar catalogs; Stellar astronomy; Stellar evolution; Exoplanet astronomy}


\section{Introduction} \label{sec:intro}

F-type stars are known to constitute the hot end of the range of stellar spectral types able to provide potentially habitable environments.  They were considered in the seminal paper by \citet{1993Icar..101..108K}, who presented first-generation climate models for Earth-type planets around main-sequence (MS) stars assuming that habitability requires, among other conditions, the presence of liquid water on the planet’s surface.  Subsequent models of stellar habitable zones (HZs) also take into account F-type stars; see, e.g., \citeauthor{2013ApJ...765..131K} (\citeyear{2013ApJ...765..131K}, \citeyear{Kopparapu_2014}), \citet{2017ARAA..55..433K}, \citet{2018Geosc...8..280R}, among a large body of other work.

More detailed studies about the suitability of F-type stars for hosting life-bearing planets include work by \citet{COCKELL1999399} and \citeauthor{Sato_Cuntz_Guerra} (\citeyear{Sato_Cuntz_Guerra}, \citeyear{Sato2017}).  Cockell’s work assumes the standard carbon-water paradigm while exploring the absorbance of DNA in the UV region of stellar spectra in order to elucidate the photobiological parameter space permitting the general possibility of life.  He found that the biochemically effective irradiance received in an F-star environment is relatively severe; that is, being 6 to 27 times higher than on the Archean Earth.  Nevertheless, according to \citet{COCKELL1999399}, a combination of UV mitigation patterns may still allow for the possibility of exolife around F-type stars even in their inner HZ regions.

\citet{Sato_Cuntz_Guerra} expanded this work toward F-type MS stars of different masses while also taking into account stellar evolutionary aspects.  Previous work by \citet{pmid11536831} assessed biologically damaging radiation for different types of stars, including F-type MS stars.  They concluded that Earth-like planets in orbit about those stars may receive less harmful UV radiation at their surfaces than Earth itself.  The relevance of UV radiation for the onset of life, especially through its ability to promote the formation of bio-macromolecules, has been discussed by, e.g., \citet{rehder}, \citet{doi:10.1021/cr400579y}, \citet{bjorn2015photobiology}, and \citet{2021uaqo.book...15G}.

Habitability in the environments of F-type stars can be established through terrestrial planets situated in stellar HZs.  Alternatively, as all currently known HZ planets orbiting F-type stars are exo-Jupiters with virtually no prospect of being habitable themselves, there is still the general possibility of habitable moons (e.g., \citealt{1997Natur.385..234W}, \citealt{2013AsBio..13...18H}, \citealt{2014AsBio..14..798H}, \citealt{2017MNRAS.472....8Z}). These studies focus on a large range of topics relevant to exomoon-based exolife, including orbital stability, tidal heating, and climate modeling.  For more recent work targeting distinct systems, see, e.g., \citet{2020AJ....159..260R} and \citet{2021PASA...38...59J}.

In summary, F-type stars have both significant advantages and disadvantages for the prospect of life.  The main disadvantage is rapid stellar evolution, noting that the stellar lifetime on the MS varies between 2 and 8 Gyr, strictly depending on the star’s mass (e.g., \citealt{1988AAS...76..411M}, \citealt{1998MNRAS.298..525P}). Notable advantages of F-type stars include, however, the broad width of their HZs (if compared to stars of lower effective temperatures) and the occurrence of UV radiation output (if occurring in moderation).  Note that the width of F-type HZs is a factor of 1.5 to 4 greater compared to the solar case, depending on the stellar mass and the adopted climate model; see \citeauthor{2013ApJ...765..131K} (\citeyear{2013ApJ...765..131K}, \citeyear{Kopparapu_2014}) and related studies.

Although the origin and sustainability of life is not explored here, the large body of existing results constitutes a strong motivation to further examine F-type star--planet systems. Here we provide detailed information on the known systems, including statistical analyses and graphical representations. In Section 2, we summarize the methods used.  Our results and discussion are given in Section 3, whereas Section 4 conveys our summary, conclusions, and outlook.

\section{Methods}

\subsection{General Approach}

To examine the statistics of F-type star--planet systems, including a preliminary evaluation of possible habitability, we utilize the NASA Exoplanet Archive\footnote{We acknowledge the heterogeneous basis of the dataset used in the NASA Exoplanet Archive.  Regrettably, this is the problem with most datasets and there is no obvious way of avoiding it. The best hope may be that more comprehensive data sets could be built in the future, which may be also less biased. A recent discussion of this issue is given in \citet{ExoPAGScienceInterestGroup:2023}; see also, e.g., \citet{Zakamska:2011} and \citet{Wright:2013}.} to collect the known data for such systems. The filtering is set to F-type single stellar systems with one planet.  We also disregard systems for which there is only limited observational data on either the planet or the star.  Hence, we are left with 206 systems fitting these specifications.

Appendix A lists the star--planet systems that form the basis of this study.  The statistics of key parameters for these systems are given in Table \ref{tab:statisticstable}. Most stars are known by their HD numbers; however, there is also a notable number of systems discovered by the {\it Kepler} mission or ones where the planet was found by CoRoT, HAT, OGLE, or WASP.  In all cases, we use the NASA Exoplanet Archive and the sources cited therein to extract the system parameters. These include the orbital parameters (semimajor axis and eccentricity), planetary radius or mass (or $M_{\rm p}\sin i$), and stellar parameters (temperature,  luminosity, and metallicity).  Of the 206 star--planet systems considered, there are 101 without published stellar luminosities.  For those stars, we adopt the TESS values as given in \citet{2019AJ....158..138S}.  As discussed in our next section, we also examine whether the stars are located on the main-sequence.

We are especially interested in determining whether a planet spends at least part, if not all, of its orbit in the stellar HZ.  The inner and outer limits of the stellar HZs are calculated using the equations given in \citet{Kopparapu_2014}; see Section 2.3 for details.

\subsection{Main-sequence Checks}
\label{sec:MS check}

Detailed checks are performed to inspect if the stars are on the main-sequence. We evaluate whether a given star is an MS star by comparing its measured values of temperature $T$ and radius $R$ with those from stellar models.  We do such a comparison using two different methods.  In the first method, which mostly serves as an introductory method for visualisation purposes, we use the $T$ and $R$ model values given in \citet{2013ApJS..208....9P} for typical F-type MS stars.  We then use the \texttt{SciPy} \citep{2020SciPy-NMeth} interpolation algorithm for linear interpolations as needed.  Next, we plug the temperature data as obtained into the interpolation function to derive the associated stellar radii.  We also consider a 10\% and 20\% cushion on either side of the respective radii and substitute the radii and temperatures into the luminosity equation $L=4\pi R^2 T^4$.  This range of cushion is chosen to account for the variation of the radius of an F-type star over the course of its main-sequence lifetime, as will be shown when comparing the interpolated models with the \texttt{MESA} models described below; see Section 3.2.  If the measured stellar luminosity falls between the minimum and maximum calculated luminosity values, the star is then considered an MS star.

In the more rigorous second method, we use the stellar evolution code \texttt{MESA}  \citep{Paxton:11,Paxton:13,Paxton:15,Paxton:18,Paxton:19,Jermyn:2022} to construct stellar models of varying mass $M$, age, and metallicity $Z$. The key parameters of the \texttt{MESA} enlist files used to build the models is the same as that used by \citet{Weinberg:2023} (see their Appendix A).  We construct models with masses in the range $M\in\{1.0, 2.0\} M_\sun$ spaced by ${\Delta}M=0.02~M_\sun$ and metallicity between about $Z_\sun/3$ to 3 $Z_\sun$ spaced by ${\Delta}Z=0.003$, where we assume solar metallicity $Z_\sun=0.02$ \citep[e.g.,][]{vonSteiger:2016}.  The metallicity parameter in \texttt{MESA} is [Fe/H] rather than $Z$; we transform from one to the other using the relations given in Appendix A of \citet{Mowlavi:2012}.

In our \texttt{MESA} models, we define the zero-age main-sequence to be when  hydrogen burning first accounts for 99.9\% of the star's total luminosity.  Similar to \citet{Bellinger:2019}, we define the terminal-age main-sequence to be when the hydrogen mass fraction at the center drops to $X=0.1$ and we consider main-sequence turn-off models down to $X = 10^{-5}$.

To determine whether a star is an MS star, we construct a rectangular error box in $T$-$R$ space that is centered on the best fit values of $T$ and $R$ and whose widths and heights are set by their $1\sigma$ error bars.  For each star, we search for stellar models with $T$-$R$ tracks that pass through the star's error box.  If any part of such a track is on the main-sequence, we refer to that star as an MS star\footnote{In the following, CoRoT-21 has been disregarded because no luminosity values are provided.  Thus, no interpolation MS checks and HZ calculations are done for this star.  CoRoT-21 has also been omitted in most figures.}.

We carry out three variations of the \texttt{MESA} MS check. In the first, we compare the data to only solar metallicity models regardless of a system’s measured metallicity.  In the second, we compare the data to all the models between  $Z_\sun/3$ and  $3Z_\sun$ regardless of a system's measured metallicity.  These first two cases mostly serve as tutorial cases, whereas the third and final case represents our main result. In the third case, we compare the data to only those models whose metallicity matches the system's measured metallicity (to within the measurement errors).  In all three tests, a star is considered MS if there is an MS model whose $T$-$R$ track passes through the systems $T$-$R$ error box.
 
In both the interpolation MS method and the \texttt{MESA} MS method, we use 100~K for the lower and upper temperature error bars if none are provided. For stellar radii, if a system only has an upper error bar, it is used for the lower error bar as well.  This procedure is also applied to the metallicity plot and the \texttt{MESA} interpolation plot.

\subsection{Governing Equations}

Next, we summarize the governing equations relevant to the planetary orbits and HZs.
The inner and outer limits of the stellar HZs are given as
\begin{equation}
    {\rm HZ}_{\rm in} = f_{\rm in}(L, T_{\rm eff})
\end{equation}
and
\begin{equation}
    {\rm HZ}_{\rm out} = f_{\rm out}(L, T_{\rm eff}) \ ,
\end{equation}
respectively, where $L$ is the stellar luminosity and $T_{\rm eff}$ is the stellar effective temperature.
Here $f_{\rm in}$ and $f_{\rm out}$ denote descriptive functions determined by the adopted
planetary climate model; they are based on Eq.~5 from \citet{Kopparapu_2014}.  For the
inner HZ limit, the runaway greenhouse concept has been employed, whereas the outer HZ limit
is based on the maximum greenhouse concept\footnote{This concept is based on the customary notion
of planetary habitability, including the relevance of aquatic environments and carbon-based chemistry.
Expansions of the concept of planetary habitability,
typically entailing larger widths of stellar HZs, have been proposed as well; see, e.g.,
review by \citet{2018Geosc...8..280R} and references therein.}. Both are calculated using the
$1 M_\oplus$ constants, however, the HZ results are not expected to be to significantly affected by this choice.

For circular orbits, the planet's distance $d$ from the respective limit of the stellar HZ is given as
\begin{equation}
   d \ = \ \begin{dcases*} \label{disteqn}
\enskip
{\rm HZ}_{\rm in} - a_{p} & \enskip \text{if~~}$a_{p} < {\rm HZ}_{\rm in}$\\
\enskip
a_{p} - {\rm HZ}_{\rm out} & \enskip \text{if~~} $a_{p} > {\rm HZ}_{\rm out}$\\
\enskip
0 & \enskip \text{otherwise}
\end{dcases*}
\end{equation}
with $a_p$ as semimajor axis (or radius $r$, in this case).

For elliptical planetary orbits, the standard case, the planet may pass through the HZ but spend only a portion of its  orbit within it.  Whether that is the case depends on the distance of the planet's pericenter and apocenter, given as 
\begin{equation} \label{zeqn1}
    r_{\rm min} = a_{p}(1-e_{p}) \ ,
\end{equation}
\begin{equation} \label{zeqn2}
    r_{\rm max} = a_{p}(1+e_{p}) \ ,
\end{equation}
where $e_p$ is the eccentricity.  Thus, for eccentric orbits we should mainly compare ${\rm HZ}_{\rm in}$  and ${\rm HZ}_{\rm out}$ to $r_{\rm min}$ and $r_{\rm max}$ instead of using Eq.~(\ref{disteqn}).

For ${\rm HZ}_{\rm in}$ and ${\rm HZ}_{\rm out}$, we also consider a possible cushion in consideration of the inherent uncertainties of those limits\footnote{See, e.g., \citet{2018Geosc...8..280R} for a review on those limits, including possible uncertainties.  For example, regarding solar-type stars, \citet{1993Icar..101..108K} introduced possible modes of CO$_2$ condensation, including one related to the maximum greenhouse effect, for defining outer HZ limits, which differ by about 0.20~au.}.

Therefore, for some of our models, we apply alterations such as
\begin{equation} \label{hzeqn1}
   {\rm HZ}_{\rm in}  \mapsto {\rm HZ}_{\rm in} - \epsilon
\end{equation}
\begin{equation} \label{hzeqn2}
   {\rm HZ}_{\rm out} \mapsto {\rm HZ}_{\rm out} + \epsilon
\end{equation}
with $\epsilon$ = 0.20 au as cushion and $\epsilon$ = 0 for no cushion.

In 42 of the 206 systems in our sample, there is only an upper limit on the eccentricity.  In these systems, the eccentricity is usually reported as a $1\sigma$, $2\sigma$, or $3\sigma$ upper limit. For the 164 other systems in our sample, the eccentricity is either measured or it is reported as zero by assumption.   For the cases where there is only an upper limit, we analyze whether the system planets are in the HZ according to four tests: in one test we assume that the eccentricity is zero and in the other three tests we assume that the eccentricity is at either the $1\sigma$, $2\sigma$, or $3\sigma$ value; for example, if the reported upper limit is $e<0.2$ at $2\sigma$, then the four cases considered are $e=0,0.1,0.2,0.3$. We find that only one of the 42 system planets is in the HZ according to these tests; it is Kepler-1708 and the planet is in the HZ based on all four tests.  This suggests that the absence of measured eccentricity values in these cases is not expected to significantly impact our HZ results.

\section{Results and Discussion}

\subsection{Stellar Parameters}

\begin{figure}[ht!]
\includegraphics[width=80mm,scale=1.5]{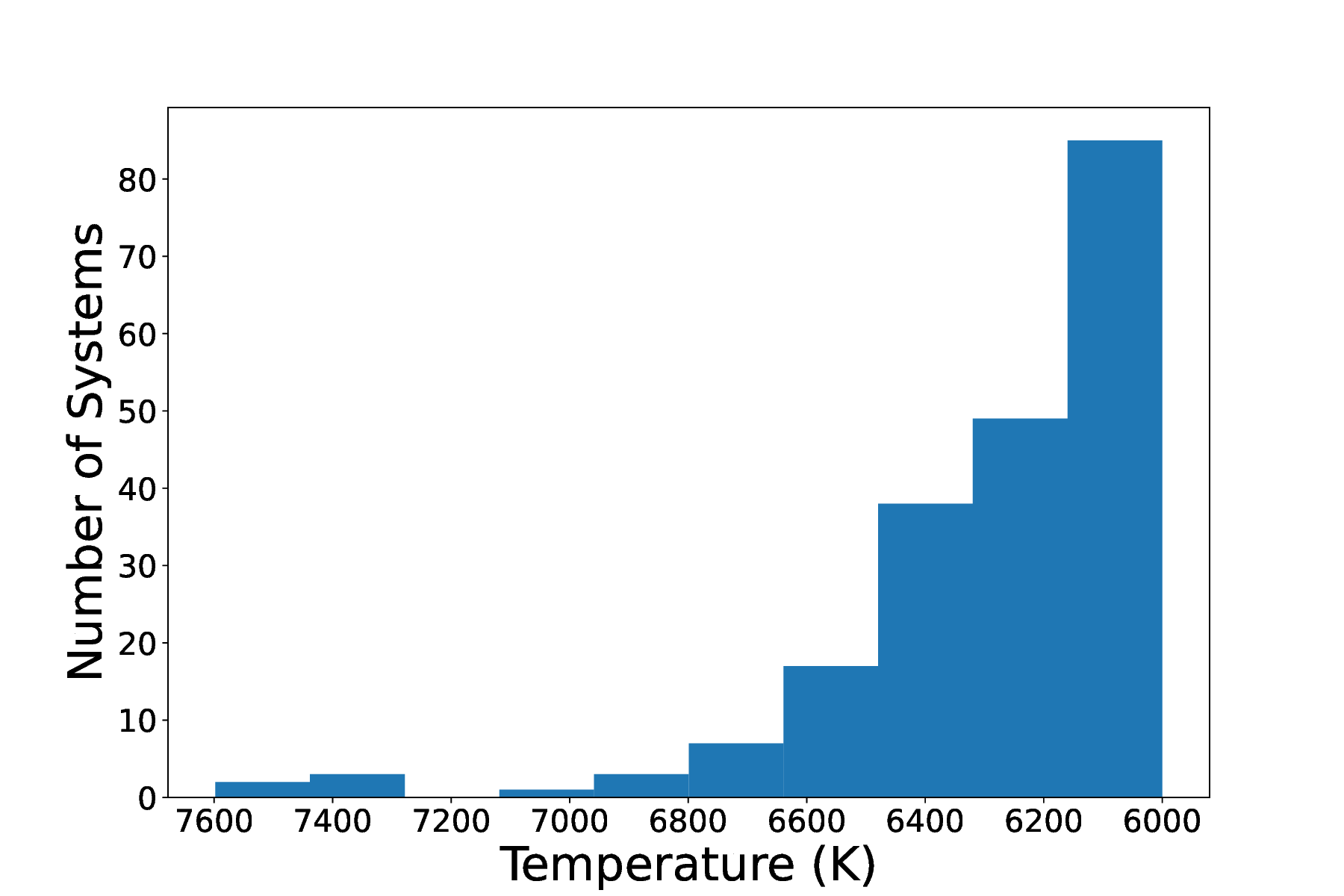}
\caption{Histogram of the stellar temperature for all systems considered.
\label{fig:fig1}}
\end{figure}

\begin{figure}[ht!]
\plotone{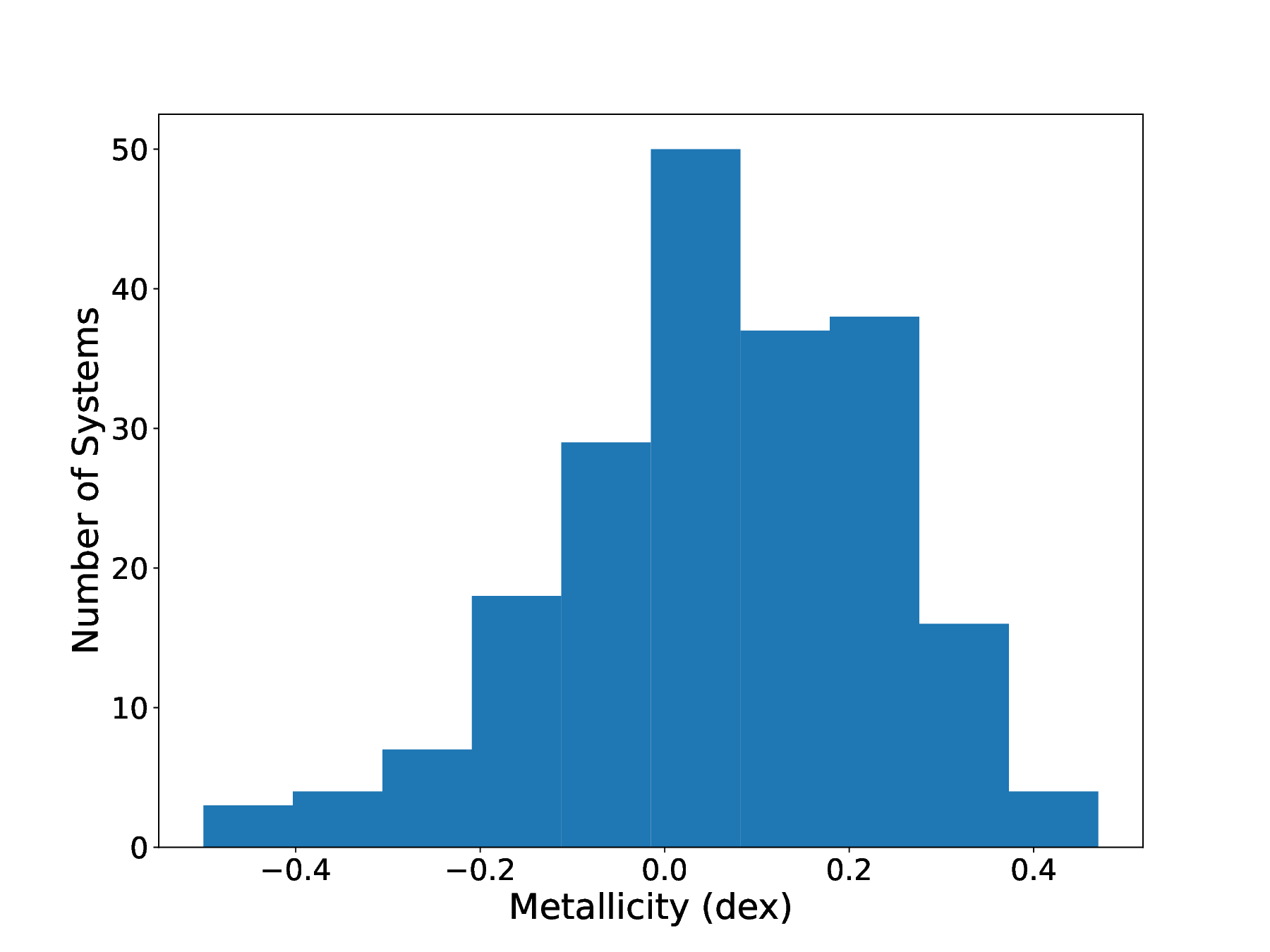}
\caption{Histogram of the stellar metallicities for all systems considered.
\label{fig:fig2}}
\end{figure}

We now present the properties of the 206 systems in our sample.
In Figures~\ref{fig:fig1} and \ref{fig:fig2},
and in Table~\ref{tab:statisticstable}, we show general properties of the stars in our sample.  Figure~\ref{fig:fig1} depicts a histogram of the stellar temperatures and Figure~\ref{fig:fig2} shows the distribution of the stellar metallicities.  The histogram of the stellar temperatures indicates that there are many more low temperature stars (in a relative sense) than high temperature stars. Only 17 stars with temperatures above 6600~K are found, whereas the vast majority of stars have
temperatures between 6000~K and 6200~K, corresponding to spectral types F8 and F9.  This result is as
expected considering the well-established frequency distribution of non-evolved stars as a function of mass, both regarding the Milky Way at large and our immediate Galactic neighborhood, as low-mass stars (or stars of low effective temperatures) are much more common than their high mass, high effective temperature counterparts \citep[e.g.,][]{2001MNRAS.322..231K,2002Sci...295...82K,2003PASP..115..763C}.

\begin{figure*}[ht!]
\plotone{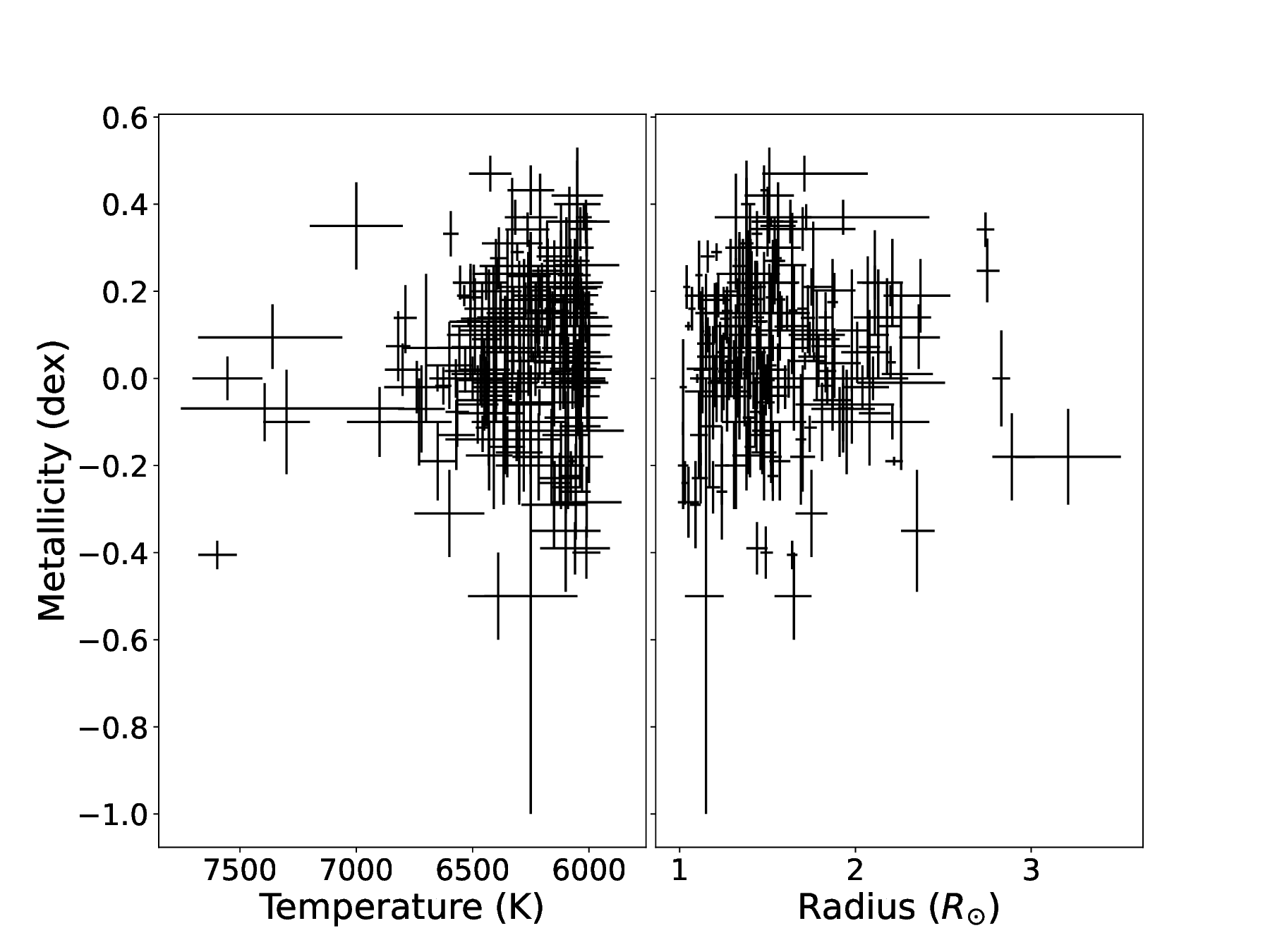}
\caption{Metallicities versus stellar temperatures (left panel) and metallicities versus stellar radii (right panel) for all systems.
\label{fig:fig3}}
\end{figure*}

Regarding stellar metallicities, both low-metallicity and high-metallicity planet-hosting stars are identified, with some preference toward the high-metallicity end.  This latter result is in agreement with previous studies \citep[e.g.,][]{1997MNRAS.285..403G,1998A&A...334..221G,2005ApJ...622.1102F,2015AJ....149...14W} indicating that metal-rich stars are more likely to harbor planets than metal-poor stars.  Figure \ref{fig:fig3} plots metallicity versus stellar temperature and radius, respectively.  The stellar metallicity values are also used in the \texttt{MESA} MS calculation.  The figure shows the range of metallicity values for the various systems,  including how they are distributed and clustered.  We see that there is a large cluster around $-$0.2 to $+$0.4 dex, further emphasizing that there is a notable tendency that planet-hosting stars are more likely to be metal-rich than metal-poor.

\subsection{Relationships between System Parameters}

\begin{figure*}[ht!]
\plotone{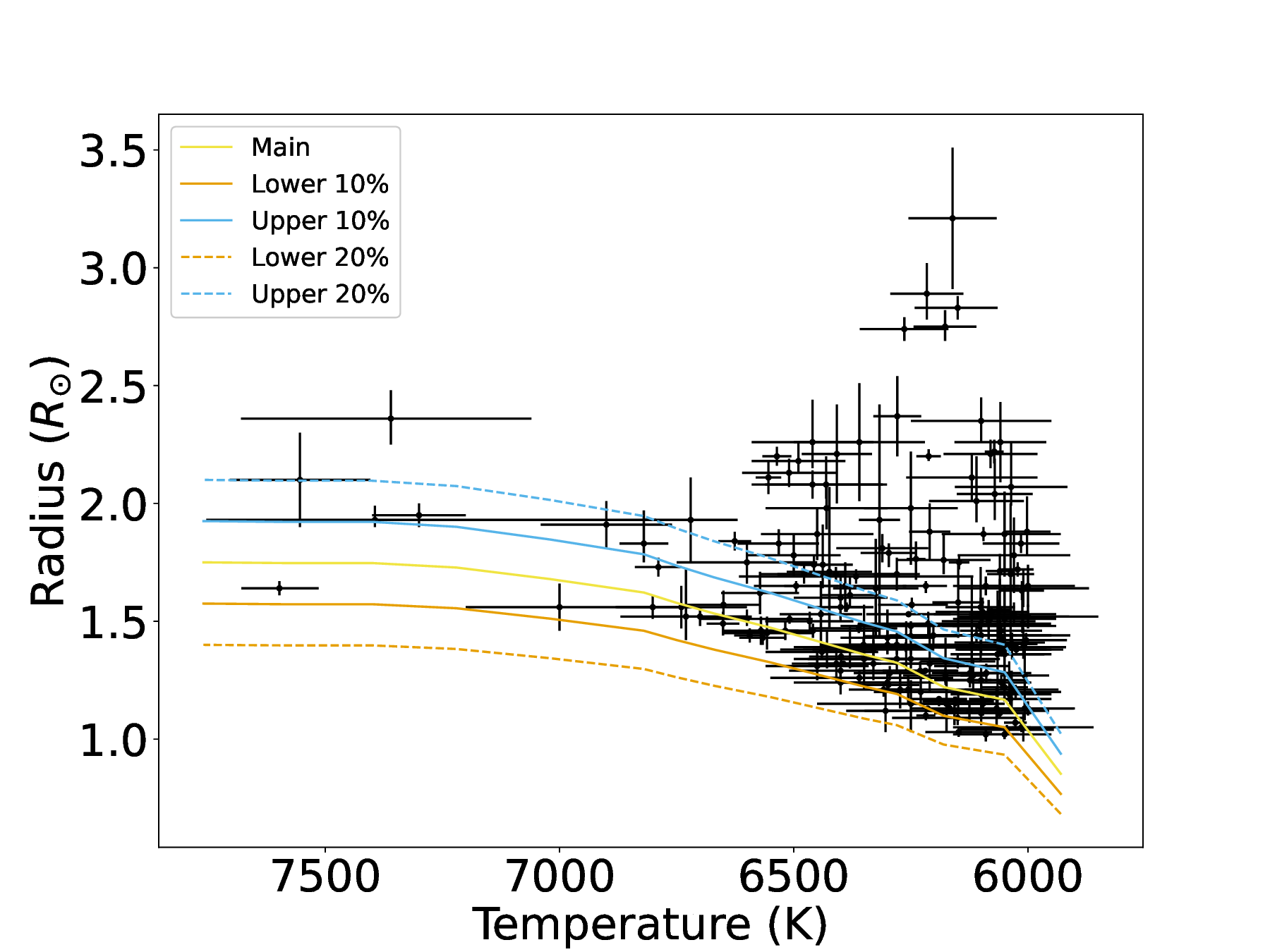}
\caption{Theoretical curves relating stellar temperatures and stellar radii for the MS F-type stars.  The main curve depicts the main relationship.  We also give upper and lower curves representing 10\% and 20\% higher and lower radii, respectively.  The actual data points from the table are overlaid.
\label{fig:fig4}}
\end{figure*}

Figure~\ref{fig:fig4} conveys the theoretical radius versus temperature curves used in our simplest, interpolation based MS check.  For these theoretical curves, we use the 10 defined temperature and radius values for F-type MS stars (F0-F9) given in \citet{2013ApJS..208....9P}.  There is one curve for the main relationship, but also upper and lower curves representing 10\% and 20\% higher and lower radii.  These upper and lower curves help give a sense of the range of temperature and radius evolution during the main sequence, and are motivated by the \texttt{MESA} MS curves discussed below.
 
Overlaid on the radius--temperature curves in Figure~\ref{fig:fig4} are the data points of the target stars. We find that the interpolation method yields 73 MS systems. There are a large number of systems with relatively low temperatures and small radii; however, there are also many systems with large radii and relatively low temperatures. These systems seem to be off the MS even when allowing for the 20\% radius range.

\begin{figure*}[ht!]
\plotone{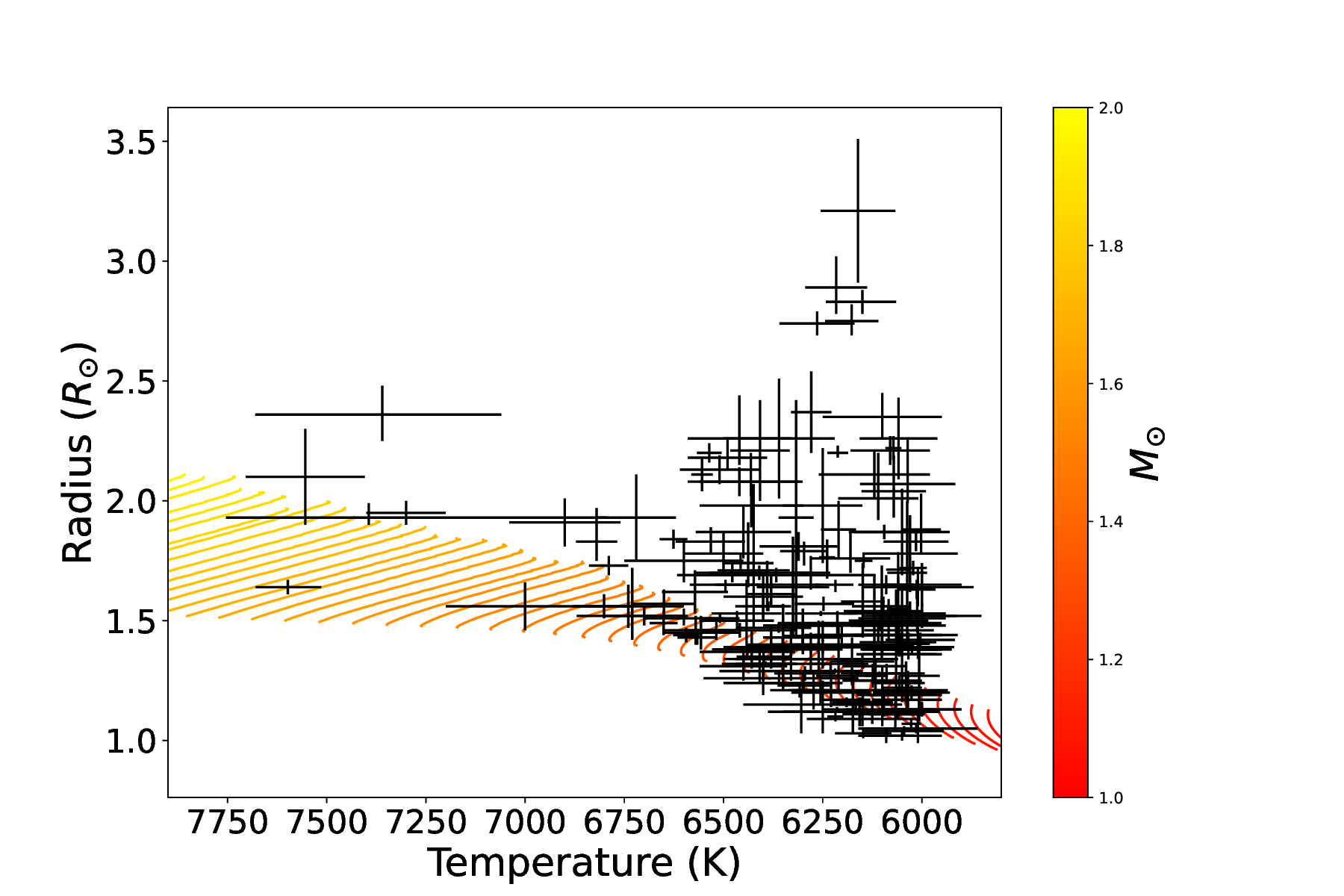}
\caption{Theoretical tracks of stellar radii versus temperatures for stars on the MS from \texttt{MESA} models of  different  masses (1.0 to 2.0~$M_\odot$, as shown in the color bar). The actual data points from the table are overlaid along with their error bars.
\label{fig:fig5}}
\end{figure*}

Next we consider the more rigorous MS check that employs \texttt{MESA}.  Figure~\ref{fig:fig5} shows the \texttt{MESA} radius versus temperature curves for 51 different stellar masses (1.0 to 2.0~$M_\odot$), with the data points overlaid.  These \texttt{MESA} runs assume solar metallicity.  The outcome indicates the large number of points concentrated in the lower temperature and lower radius regime, similar to the interpolation MS check method shown in Figure~\ref{fig:fig4}.   This check is also performed for the third dimension of metallicity, allowing us to evaluate the impact of different metallicities.

We now discuss the results of the three types of \texttt{MESA} MS checks described in Section~\ref{sec:MS check}.  In the first check, where we compare the data to only the solar metallicity models regardless of a system's measured metallicity, we find 82 MS systems. In the second check, where we  compare the data to all the models between  $Z_\sun/3$ and  $3Z_\sun$ regardless of a system's measured metallicity, we find 116 MS systems.   This check effectively provides an upper limit on the number of MS systems. In the third check, where we compare the data to only those models whose metallicity matches a system's measured metallicity, we find 58 MS systems.  Since this is our most stringent test, it is not surprising that it yields the fewest MS systems.  However, it likely underestimates the number of MS systems since there are other \texttt{MESA} parameters that are not necessarily well-constrained by observations. For example, if we allowed parameters associated with convection modeling  (e.g., mixing length theory parameters) to vary within known constraints, more of the stars might be classified as MS. However, the modeling parameter space is very large and exploring it more fully is left to future work.

Finally, we compare the interpolation and upper limit \texttt{MESA} methods to find the number of systems identified as MS systems in none, only one, or both of the methods.  We find 79 systems that are non-MS in either test, 63 that are MS in one test, and 63 that are MS in both tests.  Non-MS stars are either subgiants or giants.

\subsection{Star--Planet Systems and Habitability}

\begin{figure*}[ht!]
\plotone{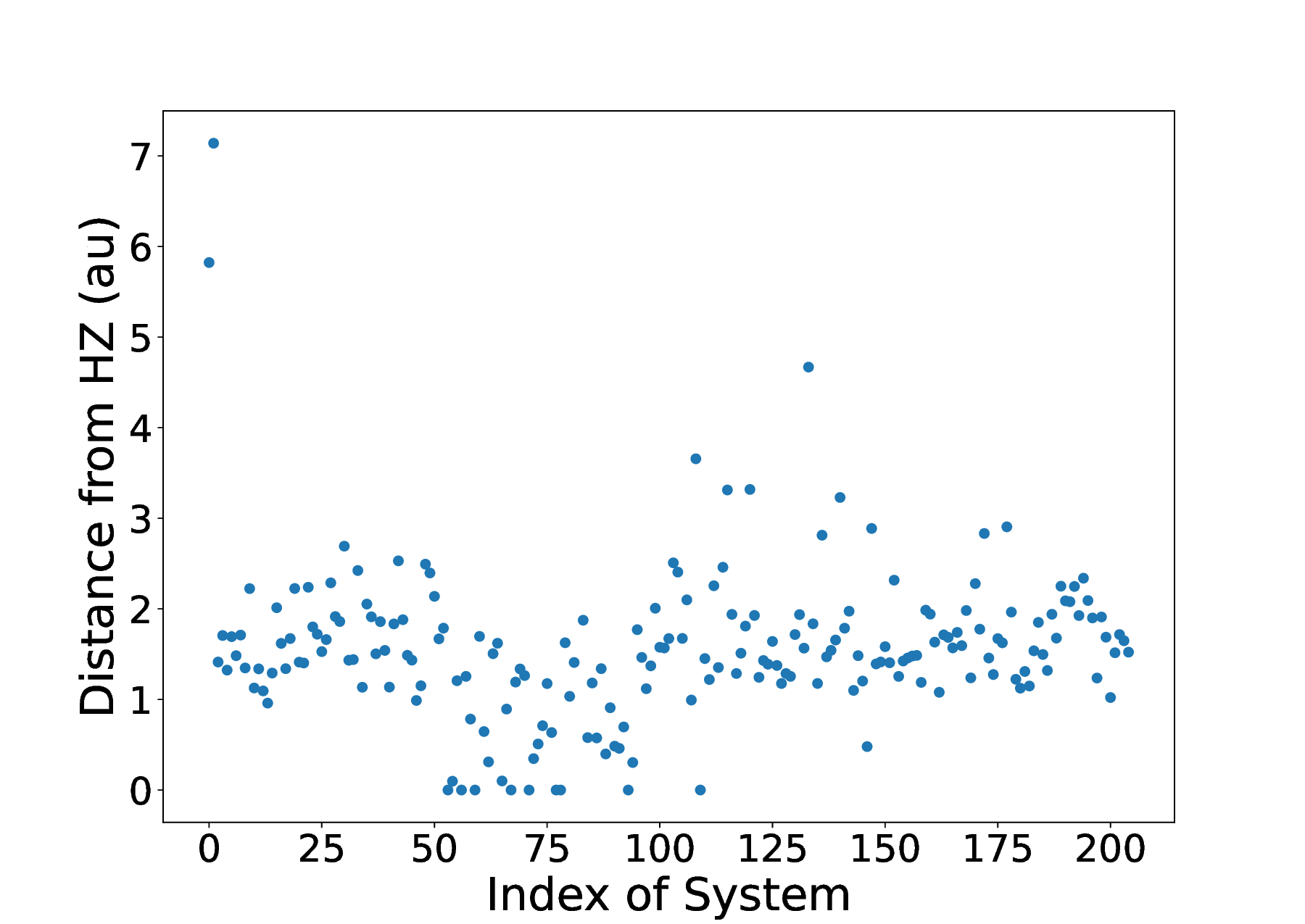}
\caption{Distance from the planet's semimajor axis to the inner or outer limit of the HZ (whichever is closer);
see Appendix~A for index information.  HD~26161 is not shown due to its outlying distance of 17.82~au.
\label{fig:fig6}}
\end{figure*}

\begin{figure}[ht!]
\includegraphics[width=80mm,scale=1.5]{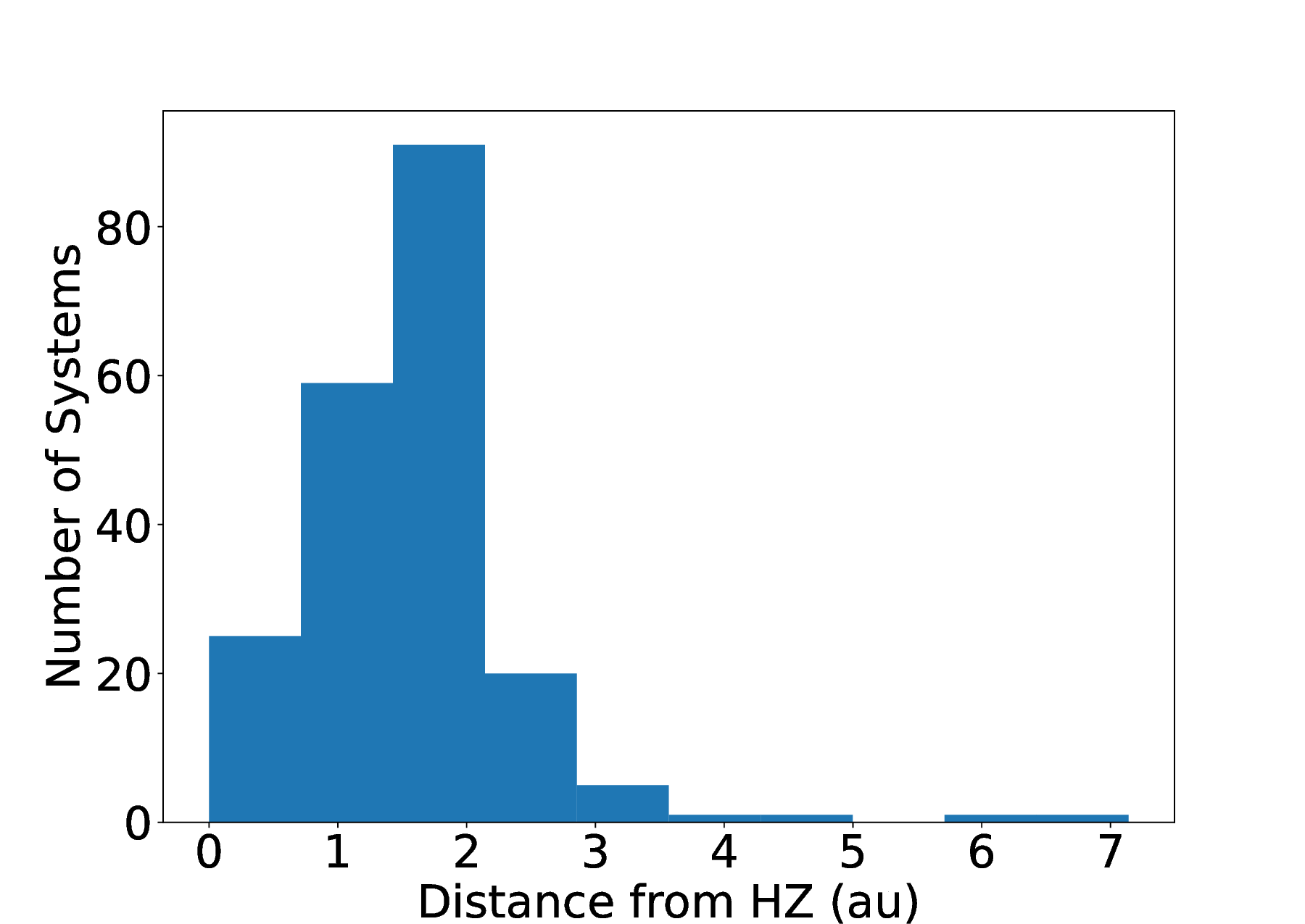}
\caption{Histogram showing the number of systems as a function of the planet's distance from the HZ.  HD~26161 is not included
due to its outlying distance of 17.82~au.
\label{fig:fig7}}
\end{figure}

A major goal of our study is to identify planets that fall within their respective HZ. As a base case, we first ignore the eccentricity of the planets and treat them as if they are on circular orbits. The results are shown in Figure \ref{fig:fig6}. The distances are calculated as the difference between the planet's semimajor axis and the inner or outer radius of the HZ (whichever is closer). We find that nine systems fall within the HZ, and two systems fall within 0.2~au of the HZ.  The vast majority of systems are between 1.0 and 2.5 au from the HZ. This is also apparent in Figure~\ref{fig:fig7} where 1 and 2 au are the largest bars in the HZ distance histogram. As noted in the caption, HD 26161 is not included in the plot as its distance of 17.82 au is an extreme outlier compared to the other points.

We then extend this testing to account for planetary eccentricity to get our main results. Equations \ref{zeqn1} and \ref{zeqn2} define the planet's periapsis and apoapsis, respectively. We can then compare these values to the HZ limits calculated previously.  We find that there are 18 systems that spend at least a part of their orbit within the HZ, and one of those systems, that is HD 111998, spends its entire orbit in the HZ. These systems are marked with the double dagger in Table \ref{tab:main_table}.  

\begin{figure*}[ht!]
\plotone{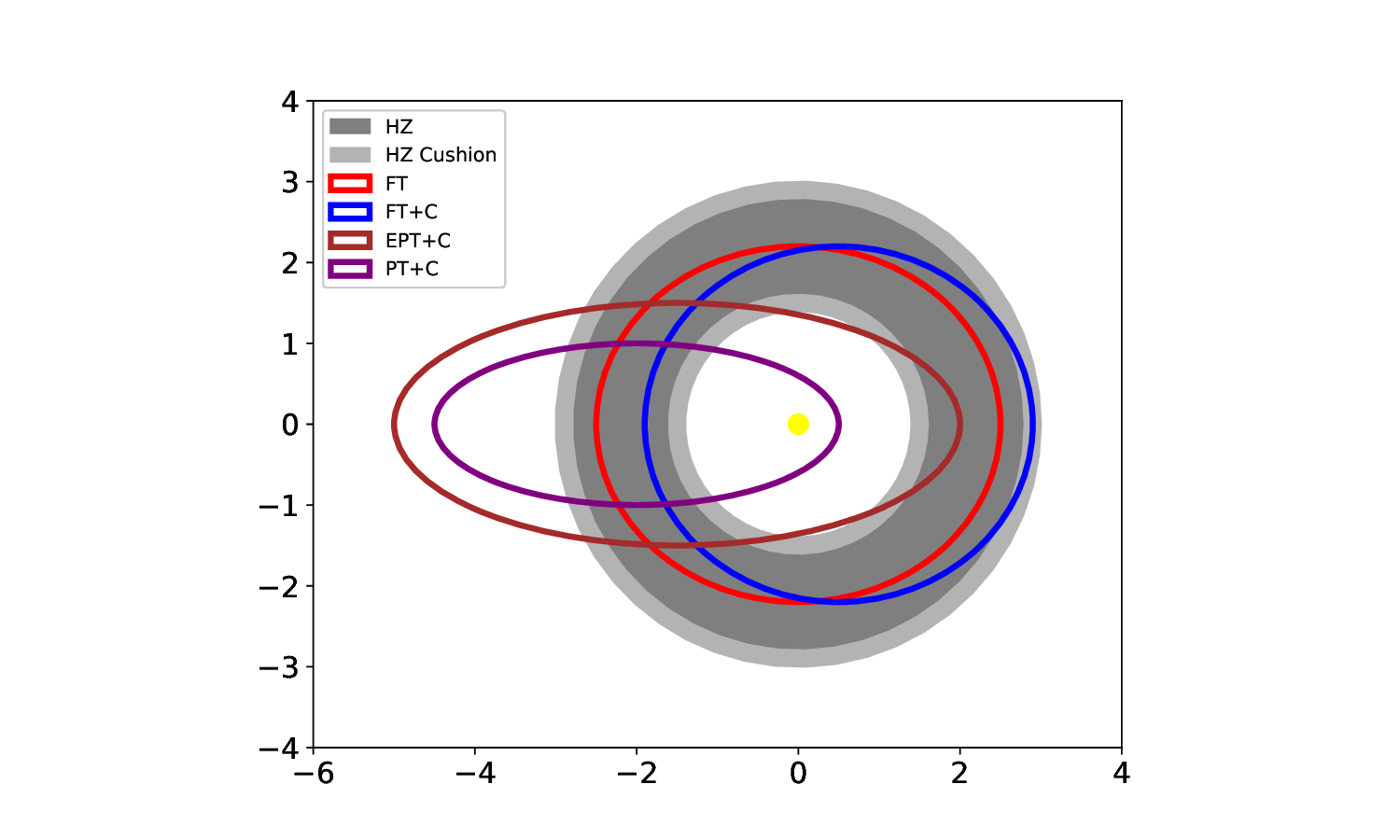}
\caption{Classification of planetary orbits relative to the stellar HZ, including the cushion; see text for details. 
\label{fig:hzcartoon}}
\end{figure*}

We break down these results into four categories: FT, FT+C, EPT+C, and PT+C. If the planet's periapsis and apoapsis are both within the HZ, we say the system is FT (or full-time). Furthermore, in another set of models, we also add a cushion to the HZ of 0.2 au in response to the inherent uncertainties when defining the HZ limits; see, e.g., \citet{2018Geosc...8..280R} and references therein. This is shown in Eqs. (\ref{hzeqn1}) and (\ref{hzeqn2}). If instead the planet's periapsis and apoapsis fall within the HZ, including the cushion, we refer to that system as FT+C. Our third case is when either the periapsis or apoapsis fall within the HZ plus cushion, which is called EPT+C or extended part-time plus cushion.  Lastly, we have the case where the planet's periapsis is interior and the apoapsis is exterior to the HZ with cushion; this case is called PT+C or part-time plus cushion. Figure~\ref{fig:hzcartoon} provides a visualization of the various types orbits as defined. 

Table \ref{tab:HZtable} shows the HZ classification  of each system as well as whether the star is defined to be on the MS by either method (interpolation or upper limit \texttt{MESA}). There are 10 MS systems where the planet is at least part-time situated in the HZ plus cushion. Table \ref{tab:stattable} provides a detailed summary of the HZ outcomes. We find that the overwhelming majority of systems are EPT+C while there seems to be a balanced mix of MS and non-MS systems.

When the upper limit eccentricity values are varied from 0 to $3\sigma$, in increments of $1\sigma$, the same 18 systems remain correlated with the HZ. The only change occurring is that Kepler-1708 changes its HZ classification; it is the only system that has an upper limit eccentricity associated with it.  Thus, Kepler-1708 could be classified either as FT (0 and $1\sigma$ test), EPT+C ($2\sigma$ test, as reported in Table~\ref{tab:stattable}), or PT+C ($3\sigma$ test).  These tests indicate that there is no need to become further engaged in statistical examinations of the upper limit eccentricity values as the HZ analysis, a key part of this study, is not significantly altered.

Furthermore, we also notice an observational bias toward short-period planetary systems.  This is evident in the available data set as 186 out of the 189 non-HZ systems are completely situated interior to the HZ, i.e., beyond the inner HZ limit.  In fact, the mean and median values for the semimajor axes of the HZ systems are considerably smaller than those of the non-HZ systems; see
Table~\ref{tab:statisticstable}.  This kind of difference is not attained for other parameters, such as the stellar and planetary masses or the stellar metallicities.

\begin{figure*}[ht!]
\plotone{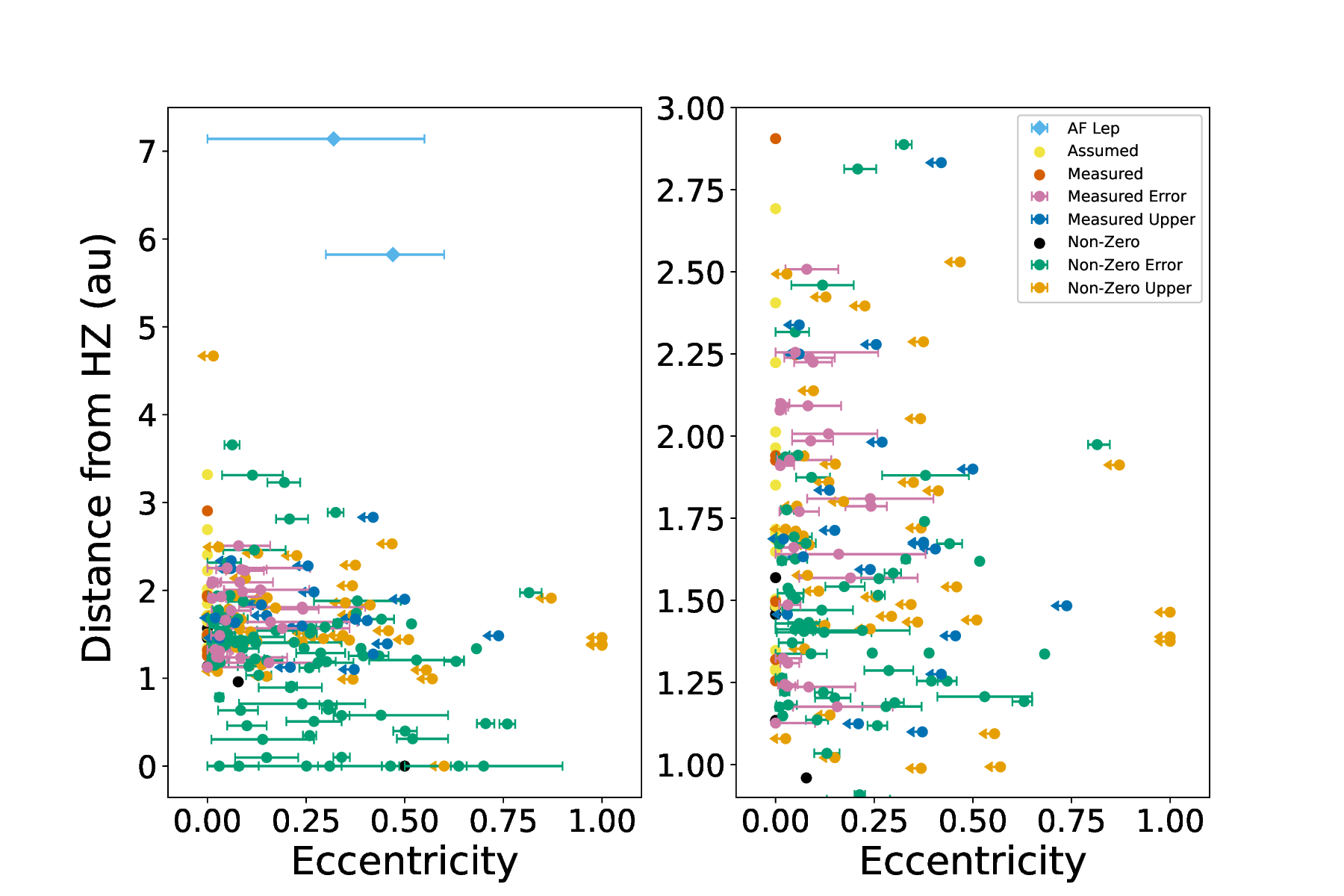}
\caption{Planetary distances (circular approximation) from the HZs in relation to the planetary eccentricities. The eccentricity is either the assumed value, an upper limit, or the actual eccentricity.  The left panel conveys the full data set, whereas the right panel conveys a zoomed-in portion. Error bars for the non-zero eccentricities are given as well.  HD~26161 is not included due to its outlying distance of 17.82~au.
\label{fig:ecc}}
\end{figure*}

Figure \ref{fig:ecc} shows the distances from the HZs (with the planets assumed in circular orbits) as a function of the planetary eccentricity. The plot can be broken down into two major categories: systems of planets with zero eccentricity and systems where the planet's eccentricity is non-zero (the standard case).  We conduct a further investigation into the zero-eccentricity systems to check if those values have been measured to be 0 (within the limits of uncertainty) or assumed to be 0. In the former case, we indicate the eccentricity value accordingly. This is broken down into Measured, Measured Error, and Measured Upper depending on the measured values provided. This includes values with no error bars, values with error bars, and upper limit values, respectively.
For the non-zero eccentricity values, we similarly break the points up into Non-Zero, Non-Zero Error, and Non-Zero Upper. We also plot AF~Lep\footnote{For AF~Lep, see Appendix A, two notably different sets of data have been given, which are subsequently referred to as AF~Lep~(1) and AF~Lep~(2).  Moreover, AF~Lep~(2) has been disregarded in the \texttt{MESA} MS checks and has been omitted in most figures due to insufficient data.} in a unique shape/color. A zoomed-in portion of the plot is given to the right of the original plot.

\begin{figure*}[ht!]
\plotone{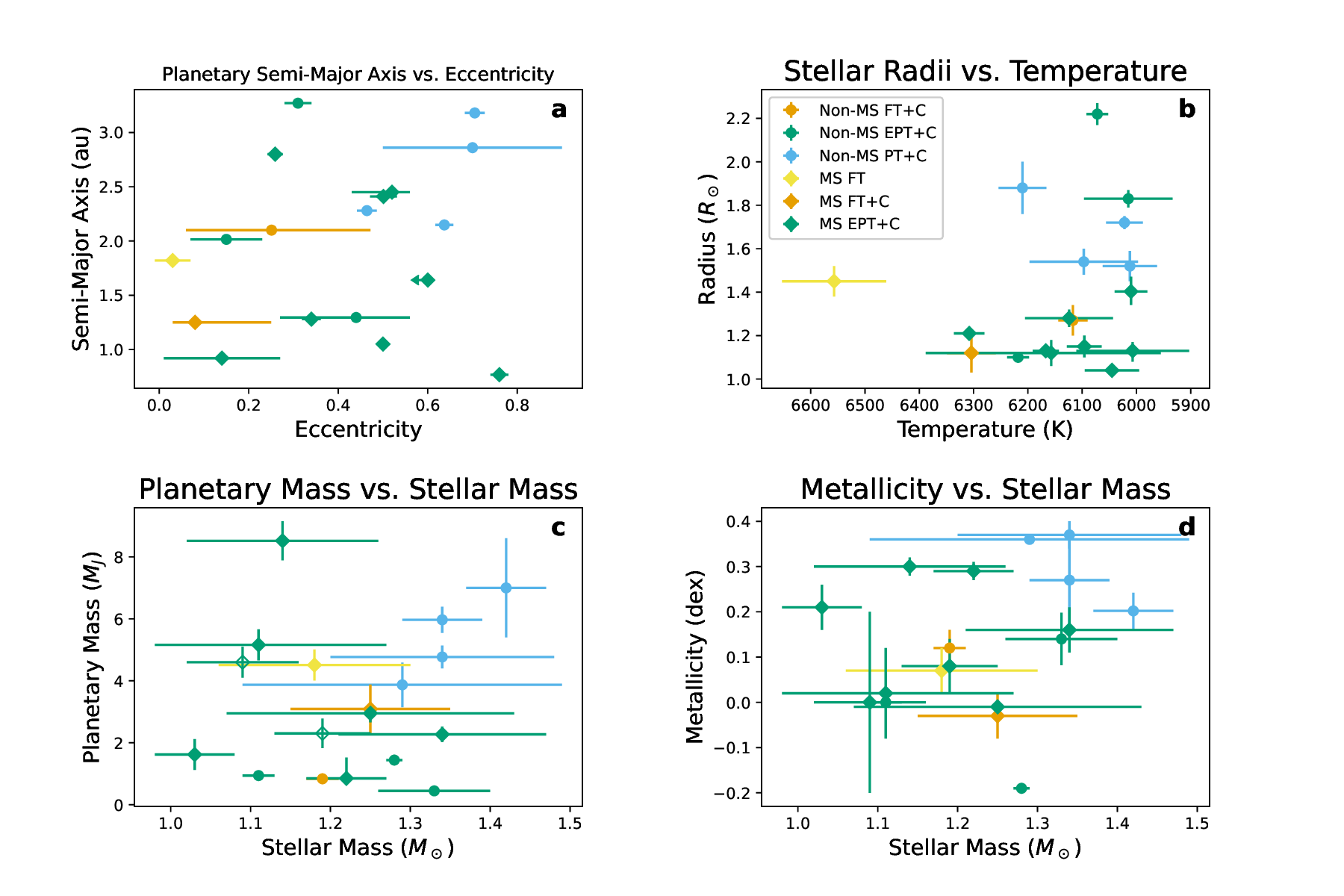}
\caption{Depiction of all 18 systems that spend at least part of their time within their respective HZs.  Empty markers in Panel c represent actual planetary mass values as opposed to $M_{\rm p}\sin i$ minimum mass values, which are represented by filled in markers.}
\label{fig:hzms}
\end{figure*}

As mentioned previously, from our main HZ case also accounting for eccentricity, we find there are 18 systems that spend at least part of their orbit in their respective HZ. Figure \ref{fig:hzms} shows the pertinent properties of these systems.  Different symbols and color coding are used for MS and non-MS systems, and the classification of the planetary orbits relative to the HZ.  The second half of Table \ref{tab:statisticstable} depicts the statistical results for key parameters for these 18 HZ systems.

Caveats and limitations of this study include: 1a) a biased dataset involving heterogeneous sources, including a bias toward short-period planets, 1b) a bias regarding the HZ systems as those have been discovered through radial velocity due to the larger planet semi-major axes, 2) MESA parameter choices leading to a potential underestimation of the number of MS systems, and 3) possible future developments concerning HZ limits as future more intricate atmospheric models might change how those limits are defined.

\section{Summary, Conclusions, and Outlook}

The aim of this study is to explore F-type stars also known to host a planet.  Although F-type stars spend considerably less time on the main-sequence compared to other types of stars, they still offer a unique set of attractive features, including the relatively large width of their HZs, thus being of general interest to astrobiology and related fields.  Here we also consider aspects of stellar evolution while making use of the stellar evolution code \texttt{MESA}. Depending on the adopted criterion, about 60 to 80 stars have been identified as MS stars.

In 18 systems out of 206 systems included in our sample, the observed planets are at least part-time located in the HZ. In one case, i.e., HD~111998 (also known as 38 Vir), the planet spends its entire orbit within the HZ.  In two other cases, i.e., HD~187085 and HD~221287, the planet also remains continuously in the HZ, if modest extensions of the HZ limits (cushions) are considered.  Note that HD~111998 and HD~221287 are MS stars, whereas HD~187085 is not.  It is noteworthy that these HZ results are not sensitive to the upper limit eccentricity values used for some systems, including systems where observationally the eccentricity is not well defined.

A focus of this work is a statistical analysis of the system properties.   As expected, most stars are of spectral type F8V and F9V, in alignment with the well-known stellar mass distribution as a function of spectral type identified in the solar neighborhood 
\citep[e.g.,][]{2001MNRAS.322..231K,2002Sci...295...82K,2003PASP..115..763C}. For stellar metallicities, there is a notable tilt toward higher than solar metallicity, a result also known for non-F-type planet-hosting stars \citep[e.g.,][and subsequent work]{1997MNRAS.285..403G,1998A&A...334..221G}.  Our study also indicates a significant number of planet-hosting non-MS stars, which are giants and subgiants.

As part of our study, we also consider cushions for both HZ limits.  This approach is informed by previous studies given by \citet{pmid21707386} and \citet{2013Icar..222....1W}.  The former work deals with climate simulations for ``land planets'' (i.e., desert worlds with limited surface water), which based on those models have a significantly extended inner HZ limit than planets with abundant surface water (akin to Earth).  Moreover, \citet{2013Icar..222....1W} continued to explore the outer limit of HZs by considering the impact of CO$_2$, including CO$_2$ clouds.  They found that in their models the outer HZ is notably extended, commensurate to the Martian orbit in the Solar System.

In our work, we also include planets that spend only part of their time within their HZs due to their orbital eccentricity.  In fact, this is the case for the majority of potentially habitable planets in our sample.  This aspect is motivated by the work of \citet{827206c76f7a4e67abd83a9972b430d2}, who examined the atmospheric conditions of bound or isolated Earth-type planets on extremely elliptical orbits in proximity of stellar HZs based on detailed general-circulation climate models.   They concluded that, in general, planetary habitability is still possible despite large variations in surface temperature as long-term climate stability merely depends on the average stellar flux received over an entire orbit, not the length of the time spent within the HZ.
Planets not located in the HZ, but near the HZ, may still be potentially habitable, especially with respect to certain types of extremophiles, as previously discussed by, e.g., \citet{pmid11234023}.  We also considered this type of augmented picture by allowing for a cushion in our analysis.

Our current study on planet-hosting F-type systems may form the basis for possible extensions in consideration of future space mission such as, e.g., LUVOIR \citep{luvo19}, HabEx \citep{gaud20}, and LIFE \citep{quan21}. Another envisioned future project is the Habitable Worlds Observatory (HWO), a large infrared / optical / ultraviolet space telescope\footnote{See {\tt https://science.nasa.gov/astrophysics/programs/ habitable-worlds-observatory} for further information.} recommended by the National Academies' Pathways to Discovery in Astronomy and Astrophysics for the 2020s.  The HWO would be the first telescope designed specifically to search for signs of life on planets outside of the Solar System; its launch is planned for the 2040s.

Our work may be useful for future studies, both regarding the possible existence of Earth-mass planets in F-type systems, as well as investigations of possibly habitable exomoons hosted by exo-Jupiters.  This latter aspect is highly relevant to F-type systems considering that the lowest-mass planet currently identified in its HZ in those systems has a mass estimate of 143 Earth masses.  Updates analyses on the survival of exomoons in different kinds of dynamic environments have been given by \citet{Dobos_2021}.

Previously, the occurrence of planets hosted by low-mass stars has been examined by \citet{2015ApJ...807...45D}, a kind of approach that deserves to be extended to stars of higher mass, albeit notably smaller datasets.  Future projects include (1) studies of planetary orbits, including cases of part-time HZ planets, (2) explorations of the relationships between planetary habitability and stellar evolution, including astrobiological aspects, and (3) assessments of exomoons for distinct systems.

\begin{acknowledgments}
    This work was supported by NSF grant No. AST-2054353.
\end{acknowledgments}

\facility {Exoplanet Archive;} This research has made use of the NASA Exoplanet Archive, which is operated by the California Institute of Technology, under contract with the National Aeronautics and Space Administration under the Exoplanet Exploration Program.

\software{\texttt{MESA} (\citealt[][]{Paxton:11, Paxton:13, Paxton:15, Paxton:18,Paxton:19, Jermyn:2022}, \url{http://mesa.sourceforge.net).}
}


\begin{deluxetable*}{lccccccc}
\caption{Statistics of Systems}
\label{tab:statisticstable}
\tablehead{\colhead{Parameter} & \colhead{Semimajor Axis} & \colhead{Planetary Mass} & \colhead{Stellar Temp.} & \colhead{Stellar Radius}& \colhead{Stellar Mass}& \colhead{Stellar Luminosity}& \colhead{Stellar Metallicity}\\
\colhead{...} & \colhead{(au)} & \colhead{($M_J$)} & \colhead{(K)} & \colhead{($R_\odot$)} & \colhead{($M_\odot$)} & \colhead{($\log~L_\odot)$}& \colhead{(dex)}}
\startdata
Mean  &  0.42  &  2.56 &  6285  &  1.56  &  1.31  &  0.517  &  0.06    \\
Median &  0.06  &  1.28 &  6219  &  1.48  &  1.29  &  0.495  &  0.06    \\
Std. Dev. &  1.63  &  3.59 &  276  &  0.38  &  0.17  &  0.235  &  0.18    \\
Minimum &  0.02 &  0.01 &  6000  &  1.02  &  1.00  &  0.060  &  $-$0.50    \\
Maximum &  20.40  &  27.20 &  7598  &  3.21  &  1.86  &  1.461  &  0.47    \\
\noalign{\smallskip}
\hline
\noalign{\smallskip}
Mean (HZ)  &  1.97  &  3.40 &  6141  &  1.40  &  1.23  &  0.363  &  0.13    \\
Median (HZ) &  2.06  &  3.02 &  6107  &  1.28  &  1.24  &  0.345  &  0.13    \\
Std. Dev. (HZ) &  0.77  &  2.32 &  141  &  0.33  &  0.11  &  0.195  &  0.15    \\
Minimum (HZ) &  0.77 &  0.45 &  6007  &  1.04  &  1.03  &  0.060  &  $-$0.19    \\
Maximum (HZ) &  3.27  & 8.52 &  6557  &  2.22  &  1.42  &  0.786  &  0.37    \\
\enddata
\tablecomments{The upper half of data corresponds to the full set of systems considered in this study, whereas the lower half of data corresponds to the 18 systems that are found to be at least partially situated in the HZ.}
\end{deluxetable*}

\begin{deluxetable*}{clcccccccl}
\caption{Assessment of Stellar Habitable Zones}
\label{tab:HZtable}
\tablehead{\colhead{Index} & \colhead{Planet} & \colhead{MS} & \colhead{Periapsis} & \colhead{Apoapsis} & \colhead{Inner HZ+C} & \colhead{Inner HZ} & \colhead{Outer HZ} & \colhead{Outer HZ+C} & \colhead{Outcome}}
\startdata
55    & HD 103891 b   & No  & 2.256  & 4.284  & 2.108   & 2.308 & 4.041 & 4.241   & EPT HZ+C  \\
56    & HD 10647 b    & No  & 1.713  & 2.317  & 0.899   & 1.099 & 1.918 & 2.118   & EPT HZ+C  \\
58    & HD 111998 b   & Yes & 1.765  & 1.875  & 1.427   & 1.627 & 2.825 & 3.025   & FT HZ    \\
61    & HD 142415 b   & Yes & 0.525  & 1.575  & 0.802   & 1.002 & 1.756 & 1.956   & EPT HZ+C \\
64    & HD 148156 b   & Yes & 1.176  & 3.724  & 1.027   & 1.227 & 2.139 & 2.339   & EPT HZ+C  \\
67    & HD 153950 b   & Yes & 0.845  & 1.715  & 1.180   & 1.380 & 2.414 & 2.614   & EPT HZ+C  \\
69    & HD 16175 b    & No  & 0.780  & 3.516  & 1.481   & 1.681 & 2.947 & 3.147   & PT HZ+C \\
73    & HD 187085 b   & No  & 1.573  & 2.627  & 1.208   & 1.408 & 2.463 & 2.663   & FT HZ+C  \\
74    & HD 191806 b   & Yes & 2.075  & 3.525  & 1.199   & 1.399 & 2.453 & 2.653   & EPT HZ+C  \\
79    & HD 221287 b   & Yes & 1.150  & 1.350  & 0.986   & 1.186 & 2.067 & 2.267   & FT HZ+C  \\
80    & HD 224538 b   & No  & 1.222  & 3.338  & 1.402   & 1.602 & 2.803 & 3.003   & PT HZ+C \\
86    & HD 31253 b    & No  & 0.725  & 1.865  & 1.674   & 1.874 & 3.285 & 3.485   & EPT HZ+C  \\
90    & HD 50554 b    & Yes & 1.203  & 3.617  & 0.947   & 1.147 & 2.012 & 2.212   & EPT HZ+C  \\
92    & HD 55696 b    & No  & 0.938  & 5.422  & 1.337   & 1.537 & 2.696 & 2.896   & PT HZ+C \\
95    & HD 86264 b    & No  & 0.858  & 4.862  & 1.780   & 1.980 & 3.457 & 3.657   & PT HZ+C \\
96    & HR 810 b      & Yes & 0.791  & 1.049  & 1.024   & 1.224 & 2.138 & 2.338   & EPT HZ+C  \\
111   & Kepler-1708 b & Yes & 0.984  & 2.296  & 0.945   & 1.145 & 2.002 & 2.202   & EPT HZ+C  \\
148   & TOI-4562 b    & Yes & 0.184  & 1.352  & 1.047   & 1.247 & 2.183 & 2.383   & EPT HZ+C 
\enddata
\tablecomments{All values in au. Cushions, see columns HZ+C, add 0.2~au buffers on either side of the HZs.
HZ limits are known to be less precise than indicated.  However, an increased level of precision has been
adopted for both numerical and tutorial reasons.}
\end{deluxetable*}

\begin{deluxetable*}{lcc}
\caption{Statistics of HZs}
\label{tab:stattable}
\tablehead{\colhead{Outcome} & \colhead{MS} & \colhead{Non-MS}}
\startdata
FT HZ    &  1  &  0   \\
FT HZ+C  &  1  &  1   \\
EPT HZ+C  &  8  &  3   \\
PT HZ+C &  0  &  4   \\
\enddata
\end{deluxetable*}


%




\appendix
\section{Summary of F-type Star--Planet Systems}

\tabletypesize{\scriptsize}
\startlongtable
\begin{longrotatetable}
\begin{deluxetable}{clccccccccccccc}
\label{tab:main_table}
\rotate
\tablecaption{F-type Star--Planet Systems}

\tablehead{\colhead{Index} & \colhead{System} & \colhead{Semimajor Axis} & \colhead{Planetary Mass} & \colhead{Eccentricity} & \colhead{Stellar Temp.} & \colhead{Stellar Radius} & \colhead{Stellar Mass} & \colhead{Stellar Luminosity} & \colhead{References}\\ 
\colhead{} & \colhead{} & \colhead{(au)} & \colhead{($M_J$)} & \colhead{} & \colhead{(K)} & \colhead{($R_\odot$)} & \colhead{($M_\odot$)} & \colhead{($\log~L_\odot)$} & \colhead{} & \colhead{}} 

\startdata
1 & AF Lep         & 7.99        & 5.237      & 0.47                & 6100     & 1.25    & 1.2      & 0.247\textsuperscript{\textdagger}      &  {\citetalias{2023AA...672A..93M}}, ' ', {\citetalias{2023ApJ...950L..19F}} \\
2 & AF Lep         & 9.3         & 4.3        & 0.32                & 6140     &  ...    & 1.13     & 0.247\textsuperscript{\textdagger}      & {\citetalias{2023AA...672A..94D}}, ' ', {\citetalias{2023ApJ...950L..19F}} \\
3 & CoRoT-1        & 0.02752     & 1.23       & \textless{}0.24     & 6298     & 1.23    & 1.22     & 0.389\textsuperscript{\textdagger}      & {\citetalias{2017AA...602A.107B}}\\
4 & CoRoT-11       & 0.0436      & 2.33       & 0                   & 6440     & 1.37    & 1.27     & 0.564\textsuperscript{\textdagger}      & {\citetalias{2010AA...524A..55G}}\\
5 & CoRoT-14       & 0.027       & 7.6        & 0                   & 6035     & 1.21    & 1.13     & 0.319\textsuperscript{\textdagger}      & {\citetalias{2011AA...528A..97T}}\\
6 & CoRoT-19       & 0.0518      & 1.11       & 0.047               & 6090     & 1.65    & 1.21     & 0.544      & {\citetalias{2012AA...537A.136G}}\\
7 & CoRoT-21       & 0.0417      & 2.26       & 0                   & 6200     & 1.95    & 1.29     &  ...       & {\citetalias{2012AA...545A...6P}}, ' ', {\citetalias{2017AA...602A.107B}} \\
8 & CoRoT-25       & 0.0578      & 0.27       & 0                   & 6040     & 1.19    & 1.09     & 0.433\textsuperscript{\textdagger}      & {\citetalias{2013AA...555A.118A}}, ' ', {\citetalias{2017AA...602A.107B}} \\
9 & CoRoT-3        & 0.0574      & 21.44      & \textless{}0.051    & 6740     & 1.56    & 1.37     & 0.589\textsuperscript{\textdagger}      & {\citetalias{2017AA...602A.107B}}\\
10 & CoRoT-35       & 0.0429      & 1.1        & 0                   & 6390     & 1.65    & 1.01     & 0.362\textsuperscript{\textdagger}      & {\citetalias{2022MNRAS.516..636S}}\\
11 & CoRoT-36       & 0.066       & 0.68       & 0                   & 6730     & 1.52    & 1.32     & 0.813\textsuperscript{\textdagger}      & {\citetalias{2022MNRAS.516..636S}}\\
12 & CoRoT-4        & 0.09        & 0.72       & 0                   & 6190     & 1.17    & 1.16     & 0.236\textsuperscript{\textdagger}      & {\citetalias{2008AA...488L..47M}}, ' ', {\citetalias{2017AA...602A.107B}} \\
13 & CoRoT-5        & 0.04947     & 0.467      & 0.09                & 6100     & 1.19    & 1        & 0.345\textsuperscript{\textdagger}      & {\citetalias{2009AA...506..281R}}\\
14 & CoRoT-6\tablenotemark{d}        & 0.0855      & 2.96       & \textless{}0.555                & 6090     & 1.02    & 1.05     & 0.204\textsuperscript{\textdagger}      & {\citetalias{2010AA...512A..14F}}\\
15 & DMPP-2         & 0.0664      & 0.437\textsuperscript{$\star$}      & 0.078               & 6500     & 1.78    & 1.44     & 0.104      & {\citetalias{2020NatAs...4..408H}}\\
16 & EPIC 211945201 & 0.148       & 0.08495    & 0                   & 6025     & 1.38    & 1.18     & 0.373\textsuperscript{\textdagger}      & {\citetalias{2018AJ....156....3C}}\\
17 & GPX-1          & 0.0338      & 19.7       & 0                   & 7000     & 1.56    & 1.68     & 0.727      & {\citetalias{2021MNRAS.505.4956B}}\\
18 & HAT-P-2        & 0.06814     & 8.62\textsuperscript{$\star$}       & 0.5172              & 6380     & 1.39    & 1.33     & 0.53       & {\citetalias{2018AJ....156..213M}}, ' ', {\citetalias{2017AA...602A.107B}} \\
19 & HAT-P-31       & 0.055       & 2.171      & 0.245               & 6065     & 1.36    & 1.22     & 0.348      & {\citetalias{2011AJ....142...95K}}, ' ', {\citetalias{2017AA...602A.107B}} \\
20 & HAT-P-34       & 0.0677      & 3.328      & 0.441               & 6442     & 1.53    & 1.39     & 0.56       & {\citetalias{2012AJ....144...19B}}, ' ', {\citetalias{2017AA...602A.107B}} \\
21 & HAT-P-40       & 0.0608      & 0.615      & 0                   & 6080     & 2.21    & 1.51     & 0.778      & {\citetalias{2012AJ....144..139H}}, ' ', {\citetalias{2017AA...602A.107B}} \\
22 & HAT-P-45       & 0.0452      & 0.892      & 0.049               & 6330     & 1.32    & 1.26     & 0.4        & {\citetalias{2014AJ....147..128H}}, ' ', {\citetalias{2017AA...602A.107B}} \\
23 & HAT-P-46       & 0.0577      & 0.493      & 0.123               & 6120     & 1.4     & 1.28     & 0.391      & {\citetalias{2014AJ....147..128H}}\\
24 & HAT-P-49       & 0.0438      & 1.73       & 0                   & 6820     & 1.83    & 1.54     & 0.814      & {\citetalias{2014AJ....147...84B}}\\
25 & HAT-P-50       & 0.0453      & 1.35       & \textless{}0.1725   & 6280     & 1.7     & 1.27     & 0.603      & {\citetalias{2015AJ....150..168H}}, ' ', {\citetalias{2017AA...602A.107B}} \\
26 & HAT-P-56       & 0.0423      & 2.18       & \textless{}0.369    & 6566     & 1.43    & 1.3      & 0.577\textsuperscript{\textdagger}      & {\citetalias{2015AJ....150...85H}}, ' ', {\citetalias{2017AA...602A.107B}} \\
27 & HAT-P-58       & 0.04994     & 0.372      & \textless{}0.1095   & 6078     & 1.53    & 1.03     & 0.456      & {\citetalias{2021AJ....162....7B}}\\
28 & HAT-P-6        & 0.05235     & 1.057      & 0                   & 6570     & 1.46    & 1.29     & 0.553      & {\citetalias{2008ApJ...673L..79N}}, ' ', {\citetalias{2017AA...602A.107B}} \\
29 & HAT-P-60       & 0.06277     & 0.574      & \textless{}0.375    & 6212     & 2.2     & 1.44     & 0.809      & {\citetalias{2021AJ....162....7B}}\\
30 & HAT-P-64       & 0.05387     & 0.58       & \textless{}0.1515   & 6457     & 1.74    & 1.3      & 0.668      & {\citetalias{2021AJ....162....7B}}\\
31 & HAT-P-66       & 0.04363     & 0.783      & \textless{}0.135    & 6002     & 1.88    & 1.25     & 0.615      & {\citetalias{2016AJ....152..182H}}\\
32 & HAT-P-69       & 0.06555     & 3.58       & 0                   & 7394     & 1.93    & 1.65     & 1          & {\citetalias{2019AJ....158..141Z}}\\
33 & HAT-P-9        & 0.05287     & 0.749      & 0.084               & 6350     & 1.34    & 1.28     & 0.418      & {\citetalias{2019AJ....157...82W}}, ' ', {\citetalias{2017AA...602A.107B}} \\
34 & HATS-11        & 0.04614     & 0.85       & \textless{}0.51     & 6060     & 1.44    & 1        & 0.403      & {\citetalias{2016AJ....152...88R}}\\
35 & HATS-12        & 0.04795     & 2.38       & \textless{}0.1275   & 6408     & 2.21    & 1.49     & 0.863      & {\citetalias{2016AJ....152...88R}}\\
36 & HATS-24        & 0.0238      & 2.26       & 0                   & 6125     & 1.12    & 1.07     & 0.19       & {\citetalias{2019RNAAS...3...35O}}\\
37 & HATS-26        & 0.04735     & 0.65       & \textless{}0.3675   & 6071     & 2.04    & 1.3      & 0.704      & {\citetalias{2016AJ....152..108E}}\\
38 & HATS-27        & 0.0611      & 0.53       & \textless{}0.8715   & 6438     & 1.74    & 1.42     & 0.669      & {\citetalias{2016AJ....152..108E}}\\
39 & HATS-3         & 0.0485      & 1.071      & 0                   & 6351     & 1.4     & 1.21     & 0.456      & {\citetalias{2013AJ....146..113B}}, ' ', {\citetalias{2017AA...602A.107B}} \\
40 & HATS-31        & 0.0478      & 0.88       & \textless{}0.3495   & 6050     & 1.87    & 1.27     & 0.619      & {\citetalias{2016AJ....152..161D}}\\
41 & HATS-35        & 0.03199     & 1.222      & \textless{}0.459    & 6300     & 1.43    & 1.32     & 0.465      & {\citetalias{2016AJ....152..161D}}\\
42 & HATS-36        & 0.05425     & 3.216      & 0.105               & 6149     & 1.16    & 1.22     & 0.215      & {\citetalias{2018AJ....155..119B}}\\
43 & HATS-39        & 0.06007     & 0.63       & \textless{}0.4125   & 6572     & 1.62    & 1.38     & 0.64       & {\citetalias{2018MNRAS.477.3406B}}\\
44 & HATS-40        & 0.04997     & 1.59       & \textless{}0.468    & 6460     & 2.26    & 1.56     & 0.903      & {\citetalias{2018MNRAS.477.3406B}}\\
45 & HATS-41        & 0.0583      & 9.7        & 0.38                & 6424     & 1.71    & 1.5      & 0.653      & {\citetalias{2018MNRAS.477.3406B}}\\
46 & HATS-42        & 0.03689     & 1.88       & \textless{}0.3435   & 6060     & 1.48    & 1.27     & 0.425      & {\citetalias{2018MNRAS.477.3406B}}\\
47 & HATS-45        & 0.05511     & 0.7        & \textless{}0.36     & 6450     & 1.31    & 1.27     & 0.425      & {\citetalias{2018AJ....155..112B}}\\
48 & HATS-52        & 0.02498     & 2.24       & \textless{}0.369    & 6010     & 1.05    & 1.11     & 0.068      & {\citetalias{2018AJ....155...79H}}, ' ',  {\citetalias{2019AJ....158..138S}}\\
49 & HATS-55        & 0.05412     & 0.921      & \textless{}0.138    & 6214     & 1.13    & 1.2      & 0.229      & {\citetalias{2019AJ....158...63E}}\\
50 & HATS-56        & 0.06043     & 0.602      & \textless{}0.0285   & 6536     & 2.2     & 1.57     & 0.898      & {\citetalias{2019AJ....158...63E}}\\
51 & HATS-64        & 0.06562     & 0.96       & \textless{}0.2265   & 6554     & 2.11    & 1.56     & 0.867      & {\citetalias{2019AJ....157...55H}}\\
52 & HATS-66        & 0.04714     & 5.33       & \textless{}0.096    & 6626     & 1.84    & 1.41     & 0.767      & {\citetalias{2019AJ....157...55H}}\\
53 & HATS-67        & 0.03032     & 1.45       & \textless{}0.0855   & 6594     & 1.44    & 1.44     & 0.547      & {\citetalias{2019AJ....157...55H}}\\
54 & HATS-68        & 0.05071     & 1.29       & \textless{}0.054    & 6147     & 1.75    & 1.35     & 0.592      & {\citetalias{2019AJ....157...55H}}\\
55 & HD 103891\textsuperscript{\textdaggerdbl}     & 3.27        & 1.44\textsuperscript{$\star$}       & 0.31                & 6072     & 2.22    & 1.28     & 0.786      & {\citetalias{2022AA...660A.124S}}\\
56 & HD 10647\textsuperscript{\textdaggerdbl}       & 2.015       & 0.94\textsuperscript{$\star$}       & 0.15                & 6218     & 1.1     & 1.11     & 0.149      & {\citetalias{2013AA...551A..90M}}\\
57 & HD 108147      & 0.102       & 0.261\textsuperscript{$\star$}      & 0.53                & 6255.73  & 1.21    & 1.22     & 0.303\textsuperscript{\textdagger}      & {\citetalias{2006ApJ...646..505B}}, {\citetalias{2019AJ....158..138S}}, {\citetalias{2008AA...487..373S}}  \\
58 & HD 111998\textsuperscript{\textdaggerdbl}      & 1.82        & 4.51\textsuperscript{$\star$}       & 0.03                & 6557     & 1.45    & 1.18     & 0.508\textsuperscript{\textdagger}      & {\citetalias{2017AA...599A..57B}}\\
59 & HD 114082      & 0.5109      & 8          & 0.395               & 6651     & 1.49    & 1.47     & 0.583      & {\citetalias{2022AA...667L..14Z}}\\
60 & HD 121504\tablenotemark{a}      & 0.33        & 1.51\textsuperscript{$\star$}       & 0.03                & 6027     & 1.07    & 1.62     & 0.15\textsuperscript{\textdagger}       & {\citetalias{2004AA...415..391M}}, {\citetalias{2019AJ....158..138S}}, ' '\\
61 & HD 142415\tablenotemark{g}\textsuperscript{\textdaggerdbl}      & 1.05        & 1.62\textsuperscript{$\star$}       & 0.5                 & 6045     &  1.04       & 1.03     & 0.06       & {\citetalias{2004AA...415..391M}}\\
62 & HD 143105\tablenotemark{h}      & 0.0379      & 1.21\textsuperscript{$\star$}       & \textless{}0.07     & 6380     & 1.611        & 1.51     & 0.554\textsuperscript{\textdagger}      & {\citetalias{2016AA...588A.145H}}\\
63 & HD 145377\tablenotemark{a}      & 0.45        & 6.02\textsuperscript{$\star$}       & 0.307               & 6014     & 1.05    & 1.2      & 0.136\textsuperscript{\textdagger}      & {\citetalias{2009AA...496..513M}}, {\citetalias{2019AJ....158..138S}}, ' '  \\
64 & HD 148156\textsuperscript{\textdaggerdbl}      & 2.45        & 0.85\textsuperscript{$\star$}       & 0.52                & 6308     & 1.21    & 1.22     & 0.25       & {\citetalias{2010AA...523A..15N}}\\
65 & HD 149026      & 0.0432      & 0.326\textsuperscript{$\star$}      & 0.051               & 6084     & 1.5     & 1.3      & 0.44       & {\citetalias{2018AJ....156..213M}}, ' ', {\citetalias{2017AA...602A.107B}} \\
66 & HD 149143\tablenotemark{a}      & 0.053       & 1.69\textsuperscript{$\star$}       & 0.016               & 6013     & 1.63    & 1.73     & 0.504\textsuperscript{\textdagger}      & {\citetalias{2017AJ....153..136S}}\\
67 & HD 153950\tablenotemark{a}\textsuperscript{\textdaggerdbl}      & 1.28        & 2.95\textsuperscript{$\star$}       & 0.34                & 6124     & 1.28    & 1.25     & 0.342\textsuperscript{\textdagger}      & {\citetalias{2009AA...496..513M}}, {\citetalias{2019AJ....158..138S}}, ' '  \\
68 & HD 155193\tablenotemark{h}      & 1.04        & 0.75\textsuperscript{$\star$}       & 0.21                & 6239     &  1.764       & 1.22     & 0.641\textsuperscript{\textdagger}      & {\citetalias{2021AA...651A..11D}}\\
69 & HD 16175\tablenotemark{g}\textsuperscript{\textdaggerdbl}       & 2.148       & 4.77\textsuperscript{$\star$}       & 0.637               & 6022     & 1.72        & 1.34     & 0.508      & {\citetalias{2016AA...591A.146D}}, ' ', {\citetalias{2023RAA....23e5022X}} \\
70 & HD 1666        & 0.94        & 6.43\textsuperscript{$\star$}       & 0.63                & 6317     & 1.93    & 1.5      & 0.73       & {\citetalias{2015ApJ...806....5H}}\\
71 & HD 17156       & 0.1614      & 3.22       & 0.6819              & 6100     & 1.44    & 1.24     & 0.412      & {\citetalias{2009AA...503..601B}}, ' ', {\citetalias{2007ApJ...669.1336F}} \\
72 & HD 179949      & 0.04439     & 0.967\textsuperscript{$\star$}      & 0.016               & 6176.2   & 1.27    & 1.22     & 0.298\textsuperscript{\textdagger}      & {\citetalias{2021ApJS..255....8R}}\\
73 & HD 187085\textsuperscript{\textdaggerdbl}      & 2.1         & 0.836\textsuperscript{$\star$}      & 0.251               & 6117     & 1.27    & 1.19     & 0.359\textsuperscript{\textdagger}      & {\citetalias{2018AA...615A.175B}}\\
74 & HD 191806\tablenotemark{h}\textsuperscript{\textdaggerdbl}      & 2.8         & 8.52\textsuperscript{$\star$}       & 0.259               & 6010     & 1.403        & 1.14     & 0.348      & {\citetalias{2016AA...591A.146D}}\\
75 & HD 205739      & 0.896       & 1.37\textsuperscript{$\star$}       & 0.27                & 6176     & 1.33    & 1.22     & 0.36       & {\citetalias{2008AJ....136.1901L}}\\
76 & HD 208487\tablenotemark{i}      & 0.524       & 0.52\textsuperscript{$\star$}       & 0.24                & 6067     & 1.13        & 1.13     & 0.242\textsuperscript{\textdagger}      & {\citetalias{2006ApJ...646..505B}}\\
77 & HD 209458      & 0.04634     & 0.665\textsuperscript{$\star$}      & 0.01                & 6026.35  & 1.2     & 1.07     & 0.231\textsuperscript{\textdagger}      & {\citetalias{2021ApJS..255....8R}}\\
78 & HD 211403      & 0.768       & 5.54\textsuperscript{$\star$}       & 0.084               & 6273     & 1.21    & 1.2      & 0.364\textsuperscript{\textdagger}      & {\citetalias{2021AA...653A..78D}}\\
79 & HD 221287\tablenotemark{i}\textsuperscript{\textdaggerdbl}      & 1.25        & 3.09\textsuperscript{$\star$}       & 0.08                & 6304     & 1.12        & 1.25     & 0.22       & {\citetalias{2007AA...470..721N}}\\
80 & HD 224538\textsuperscript{\textdaggerdbl}      & 2.28        & 5.97\textsuperscript{$\star$}       & 0.464               & 6097     & 1.54    & 1.34     & 0.47       & {\citetalias{2017MNRAS.466..443J}}\\
81 & HD 224693      & 0.192       & 0.71\textsuperscript{$\star$}       & 0.05                & 6037     & 1.7     & 1.33     & 0.577\textsuperscript{\textdagger}      & {\citetalias{2006ApJ...647..600J}}\\
82 & HD 231701      & 0.567       & 1.13\textsuperscript{$\star$}       & 0.13                & 6101     & 1.53    & 1.21     & 0.47       & {\citetalias{2018AJ....156..213M}}, ' ', {\citetalias{2007ApJ...669.1336F}} \\
83 & HD 24085\tablenotemark{h}       & 0.034       & 0.03713\textsuperscript{$\star$}    & 0.22                & 6034     & 1.404        & 1.22     & 0.376\textsuperscript{\textdagger}      & {\citetalias{2019ApJS..242...25F}}, ' ', {\citetalias{2019AJ....158..138S}} \\
84 & HD 26161       & 20.4        & 13.5\textsuperscript{$\star$}       & 0.82                & 6057.62  & 1.46    & 1.13     & 0.393\textsuperscript{\textdagger}      & {\citetalias{2021ApJS..255....8R}}\\
85 & HD 2685        & 0.0568      & 1.17       & 0.091               & 6801     & 1.56    & 1.43     & 0.668      & {\citetalias{2019AA...625A..16J}}\\
86 & HD 31253\textsuperscript{\textdaggerdbl}       & 1.295       & 0.446      & 0.44\textsuperscript{$\star$}                & 6014.94  & 1.83    & 1.33     & 0.602\textsuperscript{\textdagger}      & {\citetalias{2021ApJS..255....8R}}\\
87 & HD 332231      & 0.1436      & 0.244      & 0.032               & 6089     & 1.28    & 1.13     & 0.305      & {\citetalias{2020AJ....159..241D}}\\
88 & HD 33564\tablenotemark{j}       & 1.1         & 9.1\textsuperscript{$\star$}        & 0.34                & 6250     & 1.437        & 1.25     & 0.517\textsuperscript{\textdagger}      & {\citetalias{2005AA...444L..21G}}\\
89 & HD 35759\tablenotemark{h}       & 0.389       & 3.76\textsuperscript{$\star$}       & 0.389               & 6060     & 1.713        & 1.15     & 0.534\textsuperscript{\textdagger}      & {\citetalias{2016AA...588A.145H}}\\
90 & HD 50554\textsuperscript{\textdaggerdbl}       & 2.41        & 5.16\textsuperscript{$\star$}       & 0.501               & 6007     & 1.13    & 1.11     & 0.175\textsuperscript{\textdagger}      & {\citetalias{2003AA...410.1039P}}, {\citetalias{2019AJ....158..138S}}, {\citetalias{2019AJ....158..138S}} \\
91 & HD 52265       & 0.506       & 1.108\textsuperscript{$\star$}      & 0.213               & 6057.44  & 1.38    & 1.22     & 0.36\textsuperscript{\textdagger}       & {\citetalias{2021ApJS..255....8R}}\\
92 & HD 55696\textsuperscript{\textdaggerdbl}       & 3.18        & 3.87\textsuperscript{$\star$}       & 0.705               & 6012     & 1.52    & 1.29     & 0.43       & {\citetalias{2018AJ....156..213M}}, ' ', {\citetalias{2023RAA....23e5022X}} \\
93 & HD 75898\tablenotemark{a}       & 1.19        & 2.7\textsuperscript{$\star$}        & 0.1                 & 6061     & 1.55    & 1.43     & 0.494\textsuperscript{\textdagger}      & {\citetalias{2007ApJ...670.1391R}}, {\citetalias{2017AJ....153..136S}}, ' ' \\
94 & HD 8574        & 0.75        & 1.766\textsuperscript{$\star$}      & 0.306               & 6004.14  & 1.42    & 1.09     & 0.376\textsuperscript{\textdagger}      & {\citetalias{2021ApJS..255....8R}}\\
95 & HD 86264\textsuperscript{\textdaggerdbl}       & 2.86        & 7\textsuperscript{$\star$}          & 0.7                 & 6210     & 1.88    & 1.42     & 0.66       & {\citetalias{2009ApJ...703.1545F}}\\
96 & HR 810\textsuperscript{\textdaggerdbl}         & 0.92        & 2.27\textsuperscript{$\star$}       & 0.14                & 6167     & 1.13    & 1.34     & 0.24\textsuperscript{\textdagger}       & {\citetalias{2017AJ....153..136S}}\\
97 & K2-107         & 0.048       & 0.84       & 0                   & 6030     & 1.78    & 1.3      & 0.577\textsuperscript{\textdagger}      & {\citetalias{2017AJ....153..130E}}\\
98 & K2-167\tablenotemark{b}         & 0.091       & 0.02045    & \textless{}1        & 6011     & 1.49    & 1.01     & 0.44\textsuperscript{\textdagger}       & {\citetalias{2023AA...677A..33B}}\\
99 & K2-232         & 0.10356     & 0.398      & 0.258               & 6154     & 1.16    & 1.19     & 0.238      & {\citetalias{2018MNRAS.477.2572B}}\\
100 & K2-237         & 0.03558     & 1.363      & 0.042               & 6360     & 1.26    & 1.26     & 0.371      & {\citetalias{2020AJ....160..209I}}\\
101 & K2-260         & 0.0404      & 1.42       & 0                   & 6367     & 1.69    & 1.39     & 0.697\textsuperscript{\textdagger}      & {\citetalias{2018MNRAS.481..596J}}\\
102 & K2-34          & 0.0465      & 1.698      & \textless{}0.081    & 6149     & 1.58    & 1.23     & 0.484      & {\citetalias{2016PASP..128l4402B}}, ' ', {\citetalias{2016ApJ...825...53H}} \\
103 & K2-98          & 0.0943      & 0.10131    & 0                   & 6120     & 1.31    & 1.07     & 0.504\textsuperscript{\textdagger}      & {\citetalias{2016AJ....152..193B}}\\
104 & KELT-1         & 0.02466     & 27.23      & 0.0099              & 6518     & 1.46    & 1.32     & 0.542      & {\citetalias{2012ApJ...761..123S}}, ' ', {\citetalias{2019AJ....158..138S}} \\
105 & KELT-12        & 0.06708     & 0.95       & 0                   & 6279     & 2.37    & 1.59     & 0.892      & {\citetalias{2017AJ....153..178S}}\\
106 & KELT-21        & 0.05224     & 3.91       & 0                   & 7598     & 1.64    & 1.46     & 0.905      & {\citetalias{2018AJ....155..100J}}\\
107 & KELT-24        & 0.06969     & 5.18       & 0.077               & 6509     & 1.51    & 1.46     & 0.565      & {\citetalias{2019AJ....158..197R}}\\
108 & KELT-7         & 0.04415     & 1.28       & 0                   & 6789     & 1.73    & 1.53     & 0.758      & {\citetalias{2015AJ....150...12B}}\\
109 & Kepler-1655    & 0.1029      & 0.01699    & \textless{}0.57     & 6148     & 1.03    & 1.03     & 0.143\textsuperscript{\textdagger}      & {\citetalias{2023AA...677A..33B}}\\
110 & Kepler-1658    & 0.0544      & 5.88       & 0.0628              & 6216     & 2.89    & 1.45     & 1.206\textsuperscript{\textdagger}      & {\citetalias{2019AJ....157..192C}}\\
111 & Kepler-1708\textsuperscript{\textdaggerdbl}    & 1.64        & 4.6        & \textless{}0.6      & 6157     & 1.12    & 1.09     & 0.182      & {\citetalias{Kipping_2022}}\\
112 & Kepler-1876    & 0.0758      & 0.00755    & \textless{}0.294    & 6104     & 1.48    & 1.19     & 0.429\textsuperscript{\textdagger}      & {\citetalias{2023AA...677A..33B}}\\
113 & Kepler-39      & 0.155       & 18         & 0.121               & 6260     & 1.39    & 1.1      & 0.346\textsuperscript{\textdagger}      & {\citetalias{2011AA...533A..83B}}, ' ', {\citetalias{2015AA...575A..85B}} \\
114 & Kepler-40      & 0.08        & 2.2        & 0                   & 6510     & 2.13    & 1.48     & 0.819\textsuperscript{\textdagger}      & {\citetalias{2011AA...528A..63S}}\\
115 & Kepler-43      & 0.04436     & 3.14       & \textless{}0.0735   & 6050     & 1.38    & 1.27     & 0.349\textsuperscript{\textdagger}      & {\citetalias{2017AA...602A.107B}}, ' ', {\citetalias{2015AA...575A..85B}} \\
116 & Kepler-433     & 0.0679      & 2.82       & 0.119               & 6360     & 2.26    & 1.46     & 0.88       & {\citetalias{2015AA...575A..71A}}\\
117 & Kepler-435     & 0.0948      & 0.84       & 0.114               & 6161     & 3.21    & 1.54     & 1.129      & {\citetalias{2015AA...575A..71A}}\\
118 & Kepler-5       & 0.05064     & 2.114      & \textless{}0.072    & 6297     & 1.79    & 1.37     & 0.669      & {\citetalias{2010ApJ...713L.131K}}\\
119 & Kepler-74      & 0.084       & 0.68       & 0.287               & 6050     & 1.51    & 1.4      & 0.332\textsuperscript{\textdagger}      & {\citetalias{2013AA...554A.114H}}, ' ', {\citetalias{2015AA...575A..85B}} \\
120 & Kepler-76      & 0.0278      & 2          & \textless{}0.255    & 6409     & 1.32    & 1.2      & 0.451\textsuperscript{\textdagger}      & {\citetalias{2017AA...602A.107B}}\\
121 & Kepler-8       & 0.0483      & 0.603      & 0                   & 6213     & 1.49    & 1.21     & 0.605      & {\citetalias{2010ApJ...724.1108J}}\\
122 & MASCARA-1      & 0.043       & 3.7        & 0                   & 7554     & 2.1     & 1.72     & 1.176      & {\citetalias{2017AA...606A..73T}}\\
123 & NGTS-2         & 0.063       & 0.74       & 0                   & 6478     & 1.7     & 1.64     & 0.678\textsuperscript{\textdagger}      & {\citetalias{2018MNRAS.481.4960R}}\\
124 & NGTS-23        & 0.0504      & 0.613      & 0                   & 6057     & 1.24    & 1.01     & 0.283      & {\citetalias{2023MNRAS.518.4845J}}\\
125 & NGTS-9         & 0.058       & 2.9        & 0.06                & 6330     & 1.38    & 1.34     & 0.418\textsuperscript{\textdagger}      & {\citetalias{2020MNRAS.491.2834C}}\\
126 & OGLE-TR-132\tablenotemark{c}    & 0.03035     & 2.7        & \textless{}1        & 6210     & 1.32    & 1.305    & 0.371      & {\citetalias{2017AA...602A.107B}}, {\citetalias{2008ApJ...677.1324T}}, ' ' \\
127 & OGLE-TR-211    & 0.051       & 1.03       & 0                   & 6325     & 1.64    & 1.33     & 0.529\textsuperscript{\textdagger}      & {\citetalias{2008AA...482..299U}}, ' ', {\citetalias{2017AA...602A.107B}} \\
128 & OGLE-TR-56\tablenotemark{c}     & 0.02383     & 3.3        & \textless{}1        & 6050     & 1.36    & 1.23     & 0.35       & {\citetalias{2017AA...602A.107B}}, {\citetalias{2008ApJ...677.1324T}}, ' ' \\
129 & Pr0201\tablenotemark{e}         & 0.06        & 0.54\textsuperscript{$\star$}       & 0                   & 6174     & 1.15    & 1.24     & 0.249\textsuperscript{\textdagger}      & {\citetalias{2017AJ....153..136S}}\\
130 & Qatar-10       & 0.0286      & 0.736      & 0                   & 6124     & 1.25    & 1.16     & 0.3        & {\citetalias{2019AJ....157..224A}}\\
131 & Qatar-3        & 0.03783     & 4.31       & 0                   & 6007     & 1.27    & 1.15     & 0.279      & {\citetalias{2017AJ....153..200A}}\\
132 & Qatar-7        & 0.0352      & 1.88       & 0                   & 6387     & 1.56    & 1.41     & 0.563      & {\citetalias{2019AJ....157...74A}}\\
133 & TOI-1107       & 0.0561      & 3.35       & 0.025               & 6311     & 1.81    & 1.35     & 0.671      & {\citetalias{2022AA...664A..94P}}\\
134 & TOI-150       & 0.07037     & 2.51       & 0.262               & 6255     & 1.53    & 1.35     & 0.497      & {\citetalias{2019MNRAS.490.1094K}}, ' ', {\citetalias{2019ApJ...877L..29C}} \\
135 & TOI-1518       & 0.0389      & 2.3        & \textless{}0.015    & 7300     & 1.95    & 1.79     & 1.462\textsuperscript{\textdagger}      & {\citetalias{2021AJ....162..218C}}\\
136 & TOI-163        & 0.058       & 1.22       & 0                   & 6495     & 1.65    & 1.44     & 0.636      & {\citetalias{2019MNRAS.490.1094K}}\\
137 & TOI-201        & 0.3         & 0.42       & 0.28                & 6394     & 1.32    & 1.32     & 0.415      & {\citetalias{2021AJ....161..235H}}\\
138 & TOI-2145       & 0.1108      & 5.26       & 0.208               & 6177     & 2.75    & 1.72     & 0.997      & {\citetalias{2023MNRAS.521.2765R}}\\
139 & TOI-2154       & 0.0513      & 0.92       & 0.117               & 6280     & 1.4     & 1.23     & 0.435      & {\citetalias{2023MNRAS.521.2765R}}\\
140 & TOI-2207       & 0.0854      & 0.64       & 0.174               & 6101     & 1.56    & 1.3      & 0.484      & {\citetalias{2022AJ....164...70Y}}\\
141 & TOI-2236       & 0.05009     & 1.58       & 0                   & 6248     & 1.57    & 1.34     & 0.533      & {\citetalias{2022AJ....164...70Y}}\\
142 & TOI-2497       & 0.1166      & 4.82       & 0.195               & 7360     & 2.36    & 1.86     & 1.167      & {\citetalias{2023MNRAS.521.2765R}}\\
143 & TOI-257        & 0.1528      & 0.138      & 0                   & 6095     & 1.87    & 1.41     & 0.636      & {\citetalias{2021MNRAS.502.3704A}}\\
144 & TOI-3362       & 0.153       & 5.029      & 0.815               & 6532     & 1.83    & 1.45     & 0.739      & {\citetalias{2021ApJ...920L..16D}}\\
145 & TOI-4087       & 0.04469     & 0.73       & 0                   & 6060     & 1.11    & 1.18     & 0.176      & {\citetalias{2023ApJS..265....1Y}}\\
146 & TOI-4137       & 0.05222     & 1.44       & 0                   & 6202     & 1.44    & 1.31     & 0.439      & {\citetalias{2022AJ....164...70Y}}\\
147 & TOI-4406       & 0.201       & 0.3        & 0.15                & 6219     & 1.29    & 1.19     & 0.362      & {\citetalias{Brahm_2023}}\\
148 & TOI-4562\textsuperscript{\textdaggerdbl}       & 0.768       & 2.3        & 0.76                & 6096     & 1.15    & 1.19     & 0.253\textsuperscript{\textdagger}      & {\citetalias{2023AJ....165..121H}}\\
149 & TOI-4603       & 0.0888      & 12.89      & 0.325               & 6264     & 2.74    & 1.76     & 1.017      & {\citetalias{2023AA...672L...7K}}\\
150 & TOI-4791       & 0.0555      & 2.31       & 0                   & 6058     & 1.41    & 1.24     & 0.38       & {\citetalias{2023ApJS..265....1Y}}\\
151 & TOI-5153       & 0.158       & 3.26       & 0.091               & 6300     & 1.4     & 1.24     & 0.464\textsuperscript{\textdagger}      & {\citetalias{2022AA...666A..46U}}\\
152 & TOI-558        & 0.1291      & 3.61       & 0.298               & 6466     & 1.5     & 1.35     & 0.547      & {\citetalias{2022AJ....163....9I}}\\
153 & TOI-628        & 0.0486      & 6.33       & 0.072               & 6250     & 1.34    & 1.31     & 0.394      & {\citetalias{2021AJ....161..194R}}\\
154 & TOI-640        & 0.06608     & 0.88       & 0.05                & 6460     & 2.08    & 1.54     & 0.834      & {\citetalias{2021AJ....161..194R}}\\
155 & TOI-677        & 0.1038      & 1.236      & 0.435               & 6295     & 1.28    & 1.18     & 0.337\textsuperscript{\textdagger}      & {\citetalias{2020AJ....159..145J}}\\
156 & TOI-892        & 0.092       & 0.95       & \textless{}0.125    & 6261     & 1.39    & 1.28     & 0.431      & {\citetalias{2020AJ....160..235B}}\\
157 & WASP-101       & 0.0506      & 0.5        & 0                   & 6400     & 1.29    & 1.34     & 0.433\textsuperscript{\textdagger}      & {\citetalias{2014MNRAS.440.1982H}}, ' ', {\citetalias{2017AA...602A.107B}} \\
158 & WASP-103       & 0.01985     & 1.49       & 0                   & 6110     & 1.44    & 1.22     & 0.413      & {\citetalias{2016ApJ...823...29A}}, ' ', {\citetalias{2017AA...602A.107B}} \\
159 & WASP-106       & 0.0917      & 1.925      & 0                   & 6055     & 1.39    & 1.19     & 0.455\textsuperscript{\textdagger}      & {\citetalias{2014AA...570A..64S}}\\
160 & WASP-117       & 0.09459     & 0.2755     & 0.302               & 6038     & 1.17    & 1.13     & 0.274\textsuperscript{\textdagger}      & {\citetalias{2014AA...568A..81L}}, ' ', {\citetalias{2017AA...602A.107B}} \\
161 & WASP-118       & 0.05453     & 0.514      & 0                   & 6410     & 1.7     & 1.32     & 0.696\textsuperscript{\textdagger}      & {\citetalias{2016MNRAS.463.3276H}}\\
162 & WASP-120       & 0.0514      & 4.85       & 0.057               & 6450     & 1.87    & 1.39     & 0.678\textsuperscript{\textdagger}      & {\citetalias{2016PASP..128f4401T}}\\
163 & WASP-121       & 0.02544     & 1.183      & 0                   & 6459     & 1.46    & 1.35     & 0.519      & {\citetalias{2016MNRAS.458.4025D}}\\
164 & WASP-124       & 0.0449      & 0.6        & \textless{}0.0255   & 6050     & 1.02    & 1.07     & 0.16\textsuperscript{\textdagger}       & {\citetalias{2016AA...591A..55M}}\\
165 & WASP-131       & 0.0607      & 0.27       & 0                   & 6030     & 1.53    & 1.06     & 0.555\textsuperscript{\textdagger}      & {\citetalias{2017MNRAS.465.3693H}}\\
166 & WASP-142       & 0.0347      & 0.84       & 0                   & 6010     & 1.64    & 1.33     & 0.527\textsuperscript{\textdagger}      & {\citetalias{2017MNRAS.465.3693H}}\\
167 & WASP-15        & 0.0499      & 0.542      & 0                   & 6300     & 1.48    & 1.18     & 0.49       & {\citetalias{2009AJ....137.4834W}}, ' ', {\citetalias{2010AA...524A..25T}} \\
168 & WASP-150       & 0.0694      & 8.46       & 0.3775              & 6218     & 1.65    & 1.39     & 0.582\textsuperscript{\textdagger}      & {\citetalias{2020AJ....159..255C}}\\
169 & WASP-158       & 0.0517      & 2.79       & 0                   & 6350     & 1.39    & 1.38     & 0.507\textsuperscript{\textdagger}      & {\citetalias{2019MNRAS.482.1379H}}\\
170 & WASP-159       & 0.0538      & 0.55       & 0                   & 6120     & 2.11    & 1.41     & 0.679\textsuperscript{\textdagger}      & {\citetalias{2019MNRAS.482.1379H}}\\
171 & WASP-166       & 0.0641      & 0.101      & 0                   & 6050     & 1.22    & 1.19     & 0.288\textsuperscript{\textdagger}      & {\citetalias{2019MNRAS.482.1379H}}\\
172 & WASP-169       & 0.0681      & 0.561      & 0                   & 6110     & 2.01    & 1.34     & 0.803\textsuperscript{\textdagger}      & {\citetalias{2019MNRAS.489.2478N}}\\
173 & WASP-17        & 0.0515      & 0.486      & 0.028               & 6650     & 1.57    & 1.31     & 0.613\textsuperscript{\textdagger}      & {\citetalias{2011MNRAS.416.2108A}}, ' ', {\citetalias{2010AA...524A..25T}} \\
174 & WASP-172       & 0.0694      & 0.47       & 0                   & 6900     & 1.91    & 1.49     & 1.026\textsuperscript{\textdagger}      & {\citetalias{2019MNRAS.488.3067H}}\\
175 & WASP-174       & 0.05503     & 0.33       & 0                   & 6399     & 1.35    & 1.24     & 0.436      & {\citetalias{2020AA...633A..30M}}\\
176 & WASP-175       & 0.04403     & 0.99       & 0                   & 6229     & 1.2     & 1.21     & 0.308\textsuperscript{\textdagger}      & {\citetalias{2019MNRAS.489.2478N}}\\
177 & WASP-184       & 0.0627      & 0.57       & 0                   & 6000     & 1.65    & 1.23     & 0.534\textsuperscript{\textdagger}      & {\citetalias{2019MNRAS.490.1479H}}\\
178 & WASP-186       & 0.06        & 4.22       & 0.33                & 6361     & 1.47    & 1.22     & 0.528\textsuperscript{\textdagger}      & {\citetalias{2020MNRAS.499..428S}}\\
179 & WASP-187       & 0.0653      & 0.8        & 0                   & 6150     & 2.83    & 1.54     & 1.009\textsuperscript{\textdagger}      & {\citetalias{2020MNRAS.499..428S}}\\
180 & WASP-190       & 0.0663      & 1          & 0                   & 6400     & 1.6     & 1.35     & 0.692\textsuperscript{\textdagger}      & {\citetalias{2019AJ....157..141T}}\\
181 & WASP-22        & 0.0468      & 0.56       & 0.023               & 6000     & 1.13    & 1.1      & 0.263\textsuperscript{\textdagger}      & {\citetalias{2010AJ....140.2007M}}, ' ', {\citetalias{2017AA...602A.107B}} \\
182 & WASP-28        & 0.04469     & 0.907      & 0                   & 6150     & 1.09    & 1.02     & 0.199\textsuperscript{\textdagger}      & {\citetalias{2015AA...575A..61A}}, ' ', {\citetalias{2017AA...602A.107B}} \\
183 & WASP-31        & 0.04657     & 0.478      & 0                   & 6300     & 1.24    & 1.16     & 0.336\textsuperscript{\textdagger}      & {\citetalias{2012MNRAS.423.1503B}}, ' ', {\citetalias{2017AA...602A.107B}} \\
184 & WASP-32        & 0.0394      & 3.6        & 0.018               & 6100     & 1.11    & 1.1      & 0.21\textsuperscript{\textdagger}       & {\citetalias{2010PASP..122.1465M}}, ' ', {\citetalias{2017AA...602A.107B}} \\
185 & WASP-38        & 0.07522     & 2.691      & 0.0314              & 6150     & 1.33    & 1.2      & 0.478\textsuperscript{\textdagger}      & {\citetalias{2011AA...525A..54B}}, ' ', {\citetalias{2017AA...602A.107B}} \\
186 & WASP-48        & 0.0332      & 0.907      & 0                   & 6000     & 1.52    & 1.06     & 0.606\textsuperscript{\textdagger}      & {\citetalias{2015AA...577A..54C}}, ' ', {\citetalias{2017AA...602A.107B}} \\
187 & WASP-61        & 0.0521      & 2.06       & 0                   & 6250     & 1.39    & 1.27     & 0.449\textsuperscript{\textdagger}      & {\citetalias{2017MNRAS.464..810B}}\\
188 & WASP-62        & 0.0571      & 0.58       & 0                   & 6230     & 1.29    & 1.28     & 0.346\textsuperscript{\textdagger}      & {\citetalias{2017MNRAS.464..810B}}\\
189 & WASP-66        & 0.0546      & 2.32       & 0                   & 6600     & 1.75    & 1.3      & 0.687\textsuperscript{\textdagger}      & {\citetalias{2016ApJ...823...29A}}, ' ', {\citetalias{2017AA...602A.107B}} \\
190 & WASP-7\tablenotemark{f}         & 0.0618      & 0.96       & 0                   & 6400     & 1.24    & 1.28     & 0.557\textsuperscript{\textdagger}      & {\citetalias{2009ApJ...690L..89H}}, ' ', {\citetalias{2017AA...602A.107B}} \\
191 & WASP-71        & 0.04619     & 2.242      & 0                   & 6059     & 2.26    & 1.56     & 0.781\textsuperscript{\textdagger}      & {\citetalias{2013AA...552A.120S}}, ' ', {\citetalias{2017AA...602A.107B}} \\
192 & WASP-72        & 0.03708     & 1.5461     & 0                   & 6250     & 1.98    & 1.39     & 0.724      & {\citetalias{2013AA...552A..82G}}\\
193 & WASP-73        & 0.05512     & 1.88       & 0                   & 6036     & 2.07    & 1.34     & 0.716      & {\citetalias{2014AA...563A.143D}}, ' ', {\citetalias{2017AA...602A.107B}} \\
194 & WASP-78        & 0.0367      & 0.86       & 0                   & 6100     & 2.35    & 1.39     & 0.778\textsuperscript{\textdagger}      & {\citetalias{2017MNRAS.464..810B}}\\
195 & WASP-79        & 0.0519      & 0.85       & 0                   & 6600     & 1.51    & 1.39     & 0.679\textsuperscript{\textdagger}      & {\citetalias{2017MNRAS.464..810B}}\\
196 & WASP-82        & 0.0447      & 1.24       & 0                   & 6490     & 2.18    & 1.63     & 0.836\textsuperscript{\textdagger}      & {\citetalias{2016AA...585A.126W}}\\
197 & WASP-88        & 0.06431     & 0.56       & 0                   & 6431     & 2.08    & 1.45     & 0.746\textsuperscript{\textdagger}      & {\citetalias{2014AA...563A.143D}}, ' ', {\citetalias{2017AA...602A.107B}} \\
198 & WASP-90        & 0.0562      & 0.63       & 0                   & 6430     & 1.98    & 1.55     & 0.661\textsuperscript{\textdagger}      & {\citetalias{2016AA...585A.126W}}\\
199 & WASP-92        & 0.0348      & 0.805      & 0                   & 6280     & 1.34    & 1.19     & 0.279\textsuperscript{\textdagger}      & {\citetalias{2016MNRAS.463.3276H}}\\
200 & WASP-93        & 0.04211     & 1.47       & 0                   & 6700     & 1.52    & 1.33     & 0.673\textsuperscript{\textdagger}      & {\citetalias{2016MNRAS.463.3276H}}\\
201 & WASP-99        & 0.0717      & 2.78       & 0                   & 6180     & 1.76    & 1.48     & 0.556\textsuperscript{\textdagger}      & {\citetalias{2014MNRAS.440.1982H}}, ' ', {\citetalias{2017AA...602A.107B}} \\
202 & WTS-1          & 0.047       & 4.01       & \textless{}0.15     & 6250     & 1.15    & 1.2      & 0.126\textsuperscript{\textdagger}      & {\citetalias{2012MNRAS.427.1877C}}\\
203 & XO-3           & 0.0454      & 11.79      & 0.26                & 6429     & 1.38    & 1.21     & 0.465      & {\citetalias{2008ApJ...683.1076W}}, ' ', {\citetalias{2017AA...602A.107B}} \\
204 & XO-4           & 0.05524     & 1.612      & \textless{}0.0255   & 6400     & 1.56    & 1.32     & 0.574\textsuperscript{\textdagger}      & {\citetalias{2017AA...602A.107B}}\\
205 & XO-6          & 0.0815      & 4.4        & 0                   & 6720     & 1.93    & 1.47     & 0.568\textsuperscript{\textdagger}      & {\citetalias{2017AJ....153...94C}}\\
206 & XO-7           & 0.04421     & 0.709      & 0.038               & 6250     & 1.48    & 1.41     & 0.459\textsuperscript{\textdagger}      & {\citetalias{2020AJ....159...44C}} 
\enddata

\tablecomments{\textsuperscript{\textdagger} Luminosity values represent data points gathered by TESS (TIC).}
\tablecomments{\textsuperscript{\textdaggerdbl} Systems that spend at least part of their orbit in the HZ.}
\tablecomments{\textsuperscript{$\star$} Planetary masses that are in fact $M_{\rm p}\sin i$ minimum mass values.}
\tablecomments{Other sources for HD~149143 report much lower temperatures than the source used here. Similarly, Gaia and TESS report much lower temperatures for HD~224693, and Gaia reports a much lower temperature for OGLE-TR-56.}
\tablecomments{The upper limit eccentricity values have been scaled to be consistent with $3\sigma$. For eccentricity values that were $1\sigma$ or $2\sigma$, a normal distribution has been assumed, allowing to scale the values linearly.}
\tablecomments{The references column represent planetary, stellar, and metallicity references, respectively.  Deliberate blanks in the reference column denote a stellar or metallicity reference that matches the planetary reference for that system. If only one reference is listed, the reference is used for all three.}

\tablenotetext{a}{Using \cite{2017AJ....153..136S} for stellar parameters and planetary mass}
\tablenotetext{b}{Using \cite{2018AJ....155..136M} for planet's orbital period}
\tablenotetext{c}{Using \cite{2008ApJ...677.1324T} for all data except the planet's eccentricity and mass \citep{2017AA...602A.107B}}
\tablenotetext{d}{Using \cite{2017AA...602A.107B} for eccentricity.} 
\tablenotetext{e}{Using \cite{2012ApJ...756L..33Q} for eccentricity.} 
\tablenotetext{f}{Using \cite{Pont_2011} for eccentricity.}  
\tablenotemark{g}{Using \cite{2017AJ....153..136S} for stellar radius.}  
\tablenotemark{h}{Using \cite{2019AJ....158..138S} for stellar radius.}  
\tablenotemark{i}{Using \cite{2010ApJ...720.1290G} for stellar radius.}  
\tablenotemark{j}{Using \cite{2014MNRAS.438.2413V} for stellar radius.}

\end{deluxetable}
\end{longrotatetable}

\bibliography{ftype_filter2}{}
\bibliographystyle{aasjournal}


\end{document}